\documentclass[letterpaper,showpacs,prb,preprint]{revtex4}
\usepackage{epsfig} 
\usepackage{subfigure}
\usepackage{bm}
\setcitestyle{square}
\begin{document}
\textbf{Title:} Tuning the Catalytic Properties of Monolayer MoS$_2$ through Doping and Sulfur Vacancies
\par
\textbf{Author:} Satvik Lolla, Xuan Luo
\par
\textbf{Affiliation:} National Graphene Research and Development Center, Springfield, Virginia 22151, USA
\newpage
\textbf{Highlights:}
\par
1. The catalytic properties of TM doped MoS$_2$ with S vacancies were theoretically studied.
\par
2. Partially occupied $d$ orbitals and hybridization greatly affected the adsorption energy of O on doped MoS$_2$.
\par
3. Ir-MoS$_2$ and Rh-MoS$_2$ were the best catalysts before and after adsorption.
\par
4. Introducing S vacancies notably enhances the catalytic properties of doped MoS$_2$ sheets towards O and makes the doped MoS$_2$ a better catalyst than palladium.
\newpage
\textbf{Graphical Abstract:}
\par
\begin{figure}[htp]
	       \subfigure[]{\includegraphics[width=6cm]{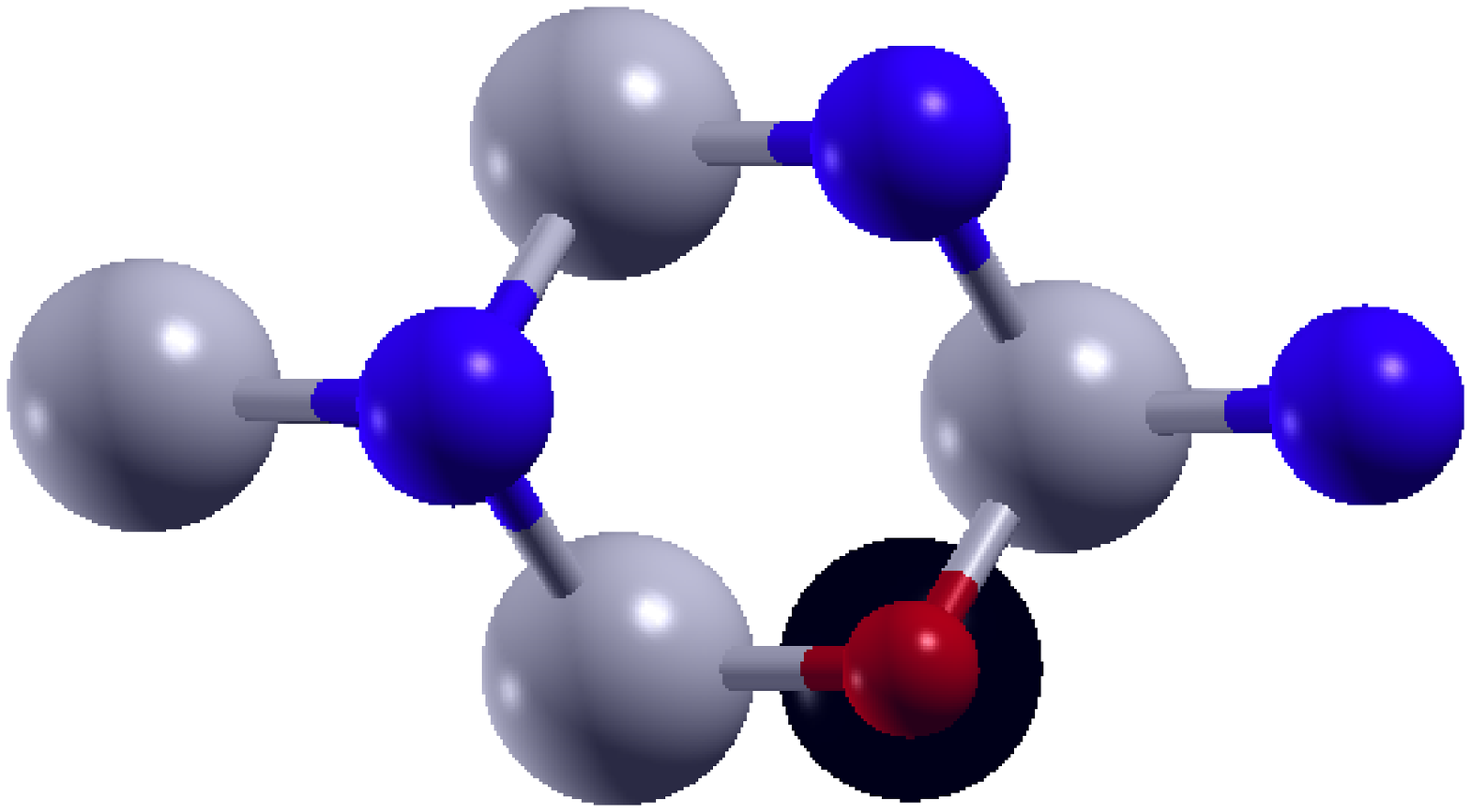}}
	       \subfigure[]{\includegraphics[width=6cm]{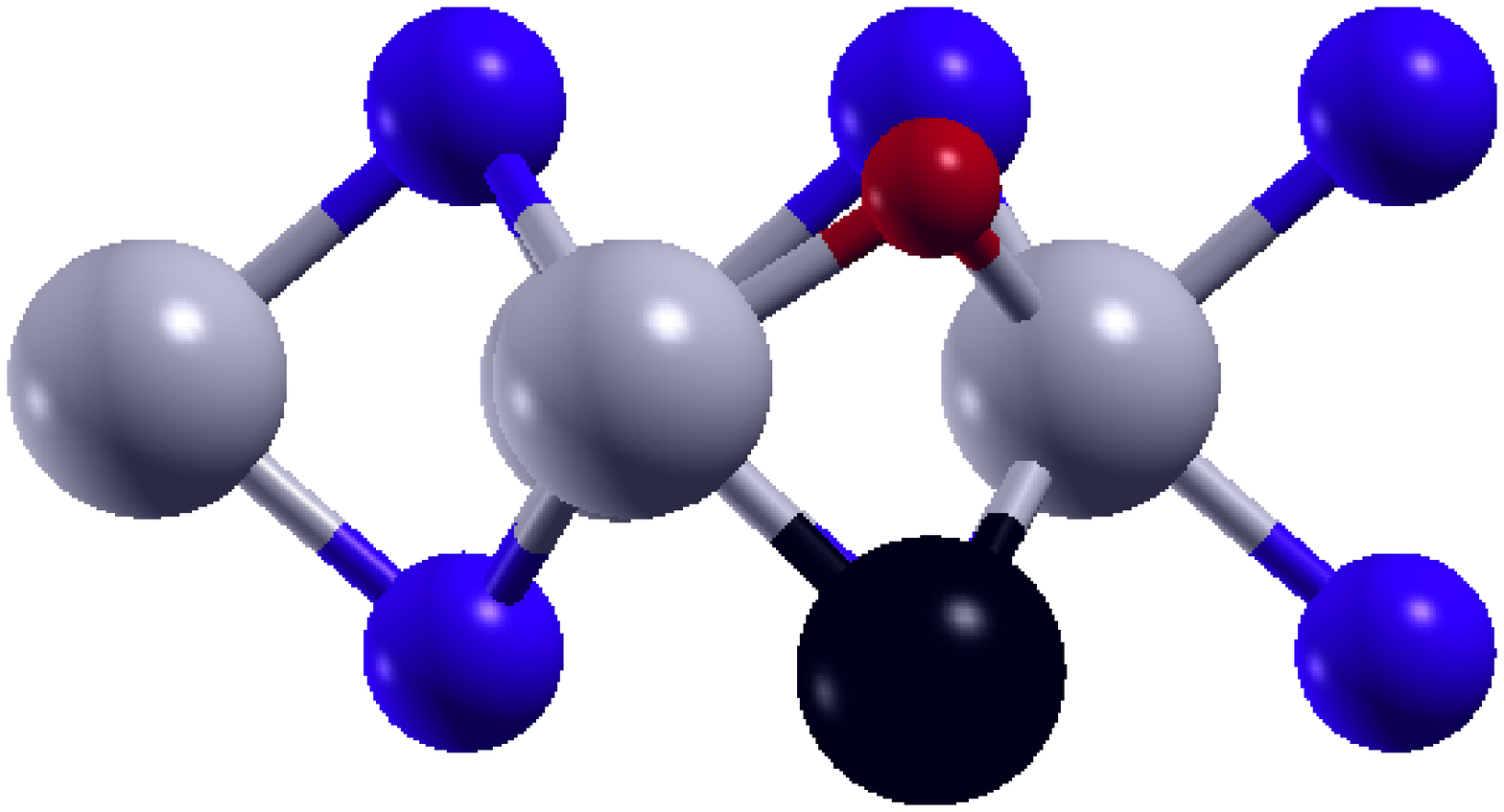}}\\
               \subfigure[]{\includegraphics[width=6cm]{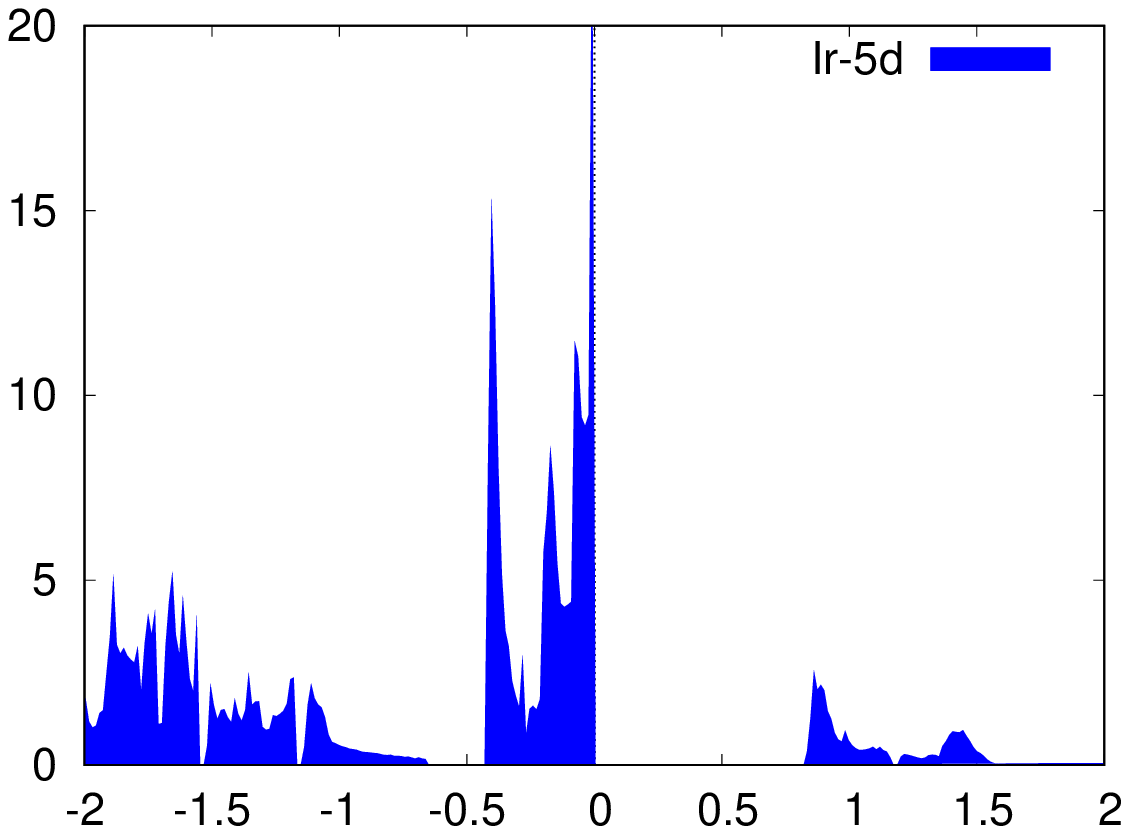}}
	       \subfigure[]{\includegraphics[width=6cm]{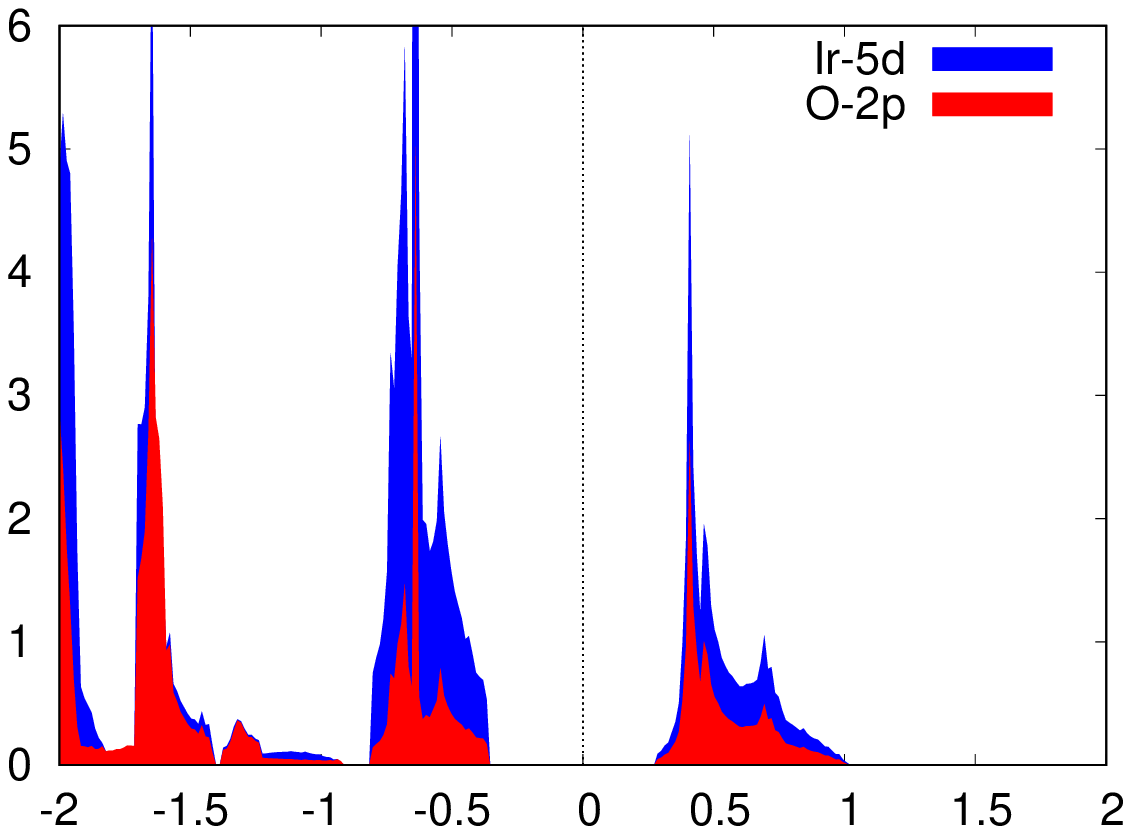}}
	\caption{Atomic structure of (a) top view of doped MoS$_2$ with sulfur vacancies after adsorption, (b) side view of doped MoS$_2$ with sulfur vacancies after adsorption. The blue, gray, black and red atoms represent the S, Mo, TM dopant, and O respectively. PDOS for (c) Ir-5d, (d) Ir-5d and O-2p after adsorption. The blue regions represent the d orbital of the dopant and the red area represents the 2p orbital of the oxygen.}
\end{figure}
A combination of substitutional doping and sulfur vacancies can modify the electronic structure	and chemical activity towards O. This may provide a route to extend the application of MoS$_2$ and other 2D monolayers to develop efficient, cheap catalysts.
\newpage
\title{Tuning the Catalytic Properties of Monolayer MoS$_2$ through Doping and Sulfur Vacancies}
\author{Satvik Lolla}
\affiliation{National Graphene Research and Development Center, Springfield, Virginia 22151, USA}
\author{Xuan Luo}
\affiliation{National Graphene Research and Development Center, Springfield, Virginia 22151, USA}
\date{\today}
\begin{abstract}
\setlength{\parindent}{1cm}
Fuel cells in vehicles are the leading cause of carbon monoxide emissions. CO is one of the most dangerous gases in the atmosphere, as it binds to the hemoglobin in blood cells 200 times easier than O$_2$. As the amount of CO in the blood stream increases, the level of oxygen decreases, which can lead to many neurological problems. To reduce the amount of CO in the atmosphere, scientists have focused on the adsorption of oxygen. The best substrates used today are platinum and palladium monolayers, which are very expensive. Because of this, researchers have searched for cheap materials, such as MoS$_2$, that are able to adsorb oxygen. However, sulfur is a chemically inert site for the oxygen, which greatly decreases the catalytic potential of monolayer MoS$_2$ sheets. Therefore, we carried out first-principles calculations to study the effect of substitutional doping and creating sulfur vacancies on the catalytic properties of MoS$_2$. We calculated the adsorption energy of O on doped MoS$_2$ sheets with vacancies, and compared it to the adsorption energy of O on a Pd monolayer. We found that doping MoS$_2$ with Ir, Rh, Co and Fe significantly decreased the adsorption energy, to below -4 eV, indicating that doped MoS2 is a more effective catalyst than Pd. Incorporating sulfur vacancies into the doped MoS$_2$ sheet was extremely effective, and decreased the adsorption energy below -6 eV. Our results show that iridium is the best catalyst as it has the lowest adsorption energy before and after sulfur vacancies were induced. We concluded that a combination of doping and creating vacancies in monolayer MoS$_2$ sheets can greatly impact the catalytic behavior and make it a more effective, less expensive catalyst than Pt and Pd.
\end{abstract}
\newpage
\maketitle
\section{Introduction}
Climate change drastically affects the air we breathe. Various gases, such as CO$_2$, CH$_4$, NO$_2$ and NO are produced\cite{searchinger2008use}; however, the most dangerous greenhouse gas by far is CO, which may be the cause of more than one-half of the fatal poisonings reported in many countries \cite{ernst1998carbon}. The hemoglobin in blood cells binds to carbon monoxide 200 times easier than oxygen does, which reduces the amount of oxygen in the blood while simultaneously increasing the level of carbon monoxide in the body\cite{raub2000carbon}. This deprives the heart, brain, and other vital organs of oxygen. It can cause significant medical issues, including memory loss, seizures and neurological impairment\cite{kao2005carbon}. In order to limit the effects of carbon monoxide, scientists have focused on fuel cells, as they are the leading cause of CO in the atmosphere\cite{weaver2009carbon}. Carbon monoxide can be created when fuel does not burn properly, or when an automobile engine is run in an enclosed space. To stop the carbon monoxide before it exits the vehicle, researchers use adsorption, the process in which a solid holds molecules of a gas or liquid. Many common adsorbents include clay\cite{gaines1953adsorption}, colloids\cite{deshmukh2015hard} and nanoparticles\cite{2004colloidal}. Nanoparticles are often used as catalysts in fuel cells to reduce carbon monoxide emissions.
\par
The two nanoparticle catalysts used to adsorb oxygen today are platinum and palladium\cite{li2019mechanism, gauthard2003palladium}. These two precious metals are extremely expensive and rare. However, they are very effective as they can easily break the carbon-oxygen bond in CO\cite{zhou2010enhancement}. In the hopes of developing new, promising catalysts, past research has focused on doping\cite{zhang2009study} and using other transition metals. The catalytic properties of palladium doped with gold and platinum doped with gold have been studied, but these mixtures have proven to be less effective as the pure metal catalysts\cite{gao2012pd, liu2013defective}. Another alloy that researchers have studied is a palladium nickel alloy\cite{lim2013density}. The Pd-Ni alloy was only an effective catalyst if the oxygen molecules were adsorbed by the palladium\cite{phillips2018chemoselective}. Silver is also being studied as a possible nanoparticle catalyst; however, silver has a very high adsorption energy, so it is not a good catalyst\cite{wu2017activity}. Emerging research details the use of a silver-copper mixture, usually Ag$_{12}$Cu, but this alloy is not as effective as Pt or Pd\cite{zhang2017activity, li2013density}.
\par
Because researchers have not been able to find a new catalyst by doping platinum and palladium, they have investigated the catalytic properties of two-dimensional (2D) monolayers.
\par
Some of these new 2D materials were classified as Transition Metal Dichalcogenides (TMDs). TMDs can be represented as MX$_2$ where M represents molybdenum or tungsten, and X represents sulfur, selenium, or tellerium. These TMDs are highly reactive due to their buckling height\cite{chhowalla2013chemistry}. TMDs have been studied as catalysts for fuel cells\cite{2005biomimetic}, but experiments have shown that the X atom is not an active adsorption site, which means that the MX$_2$ is highly unstable after the oxygen is adsorbed on the X atom\cite{wang2015transition}. In order to make the MX$_2$ more reactive, the TMD could be improved by using substitutional doping with transition metals, such as cobalt, nickel, zinc or copper\cite{elizondoco}.
\par
We chose MoS$_2$ as our TMD because of its high reactivity and low cost. We are currently unaware of studies that have used both doping and vacancies to adsorb O. Therefore, our objective is to determine whether the catalytic properties of MoS$_2$ can be improved through a combination of substitutional doping and sulfur vacancies. Specifically, we doped with Iron (Fe), Cobalt (Co), Nickel (Ni), Copper (Cu), Rhodium (Rh), Silver (Ag), and Iridium (Ir). We calculate the adsorption energy of O and the Projected Density of States for each compound.
\par
In Section II, we detailed our methods to perform first-principle calculations. In Sec. III, we present our results on the adsorption energies. In Sec. IV, we discuss and compare our results with experimental and other theoretical research. Finally, our conclusion and future work are found in Sec. V.
\section{Methods}
\subsection{Computational Methods}
We conducted first-principles calculations based on density functional theory (DFT) using the ABINIT\cite{gonze2016} code. We used pseudopotentials in projector augmented-wave (PAW) approximations\cite{blochl1994projector} and exchange correlation functionals using the Generalized Gradient Approximation in Perdew-Burke-Ernzehorf approximation (GGA-PBE)\cite{perdew1996generalized}. These projectors were compiled using the AtomPAW software\cite{holzwarth2001projector}. The electron configuration and radius cutoff to generate the PAW pseudopotentials are shown in Table \ref{valence}.
\subsection{Convergence and Relaxation}
The kinetic energy cutoff and Monkhorst-Pack k-point grids were converged for all materials. The self-consistent field (SCF) total energy tolerance was set as 1.0 $\times$ $10^{-10}$ Hartree. Once this tolerance was reached twice consecutively, the SCF iterations were terminated. The kinetic energy cutoff and Monkhorst-Pack k-point grids were considered converged when the differences in total energies were less than 1.0 $\times$ $10^{-4}$ Ha (~ 3 meV) twice consecutively. With the converged values, we performed a structural relaxation using the Broyden-Fletcher-Goldfarb-Shanno (BFGS) algorithm to determine the atomic positions and lattice constants of the material. The relaxation finished when the maximum absolute force on each atom was less than 5.0 $\times 10^{-5}$ Hartree/Bohr. The maximum dilation of the atomic positions from the initial position with each SCF step was 1.05. All structures were fully relaxed. Using these converged values, we calculated the total energy of the system.
\subsection{Materials}
We used the hexagonal monolayer crystal structure of MoS$_2$ in this study. To adsorb oxygen onto these materials, we used a 2$\times$2 supercell and we placed the oxygen in the middle of the crystal, on the molybdenum atom, and on the sulfur atom. The converged kinetic energy cutoff and k-point mesh was 29 Hartree and $6\times6\times1$ respectively.
\par
A palladium monolayer was also used as a catalyst because many automobiles today use Pd-based catalysts. We wanted to compare the adsorption energies of O on Pd to the adsorption energies of O on the doped MoS$_2$ to examine the efficiency of the MoS$_2$. The palladium monolayer discovered by Shah et al.\cite{shah2017oxygen} was used to adsorb oxygen. The hexagonal monolayer crystal structure shown in Figure \ref{Pdsites} was used. Oxygen was placed on top of the Pd atom and at the center of the hexagonal monolayer 2$\times$2 cell. The converged kinetic energy cutoff and k-point mesh was 29 and $2\times2\times1$ respectively.
\par
We used Fe, Ni, Co, Cu, Ag, Ir and Rh as our dopants. To dope the MoS$_2$, we used a 2$\times$2 cell with 12 atoms. First, the plane wave kinetic energy cutoff of the dopant was converged. These converged cutoffs are shown in Table \ref{dopant ecut}. We compared this to the energy cutoff of the MoS$_2$ and used the larger value in our total energy calculations. One sulfur atom was then replaced with a transition metal, and the system was fully relaxed. We added an oxygen atom on top of the transition metal and then relaxed the lattice constants. The doping percentage was increased from 8.3\% to 9.1\% by introducing a sulfur vacancy. In order to create sulfur vacancies in the doped MoS$_2$, we removed the sulfur atom closest to the dopant. After this structure was fully relaxed, we added an oxygen atom to the same location where the sulfur atom was before.
\subsection{Adsorption Energy}
We calculated the adsorption energy of our materials on oxygen. We calculated the total energy of the cell, and then recalculated it after placing an oxygen atom inside the crystal. Adsorption energies were calculated using the following formula:
\begin{equation}
	E_{adsorption} = E_{monolayer + O} - E_{monolayer} - E_{O}
\end{equation}
where $E_{adsorption}$, $E_{monolayer + O}$, $E_{monolayer}$, and $E_{O}$ represent the adsorption energy, energy of the monolayer MoS$_2$ after adsorption, the energy of the monolayer MoS$_2$ before adsorption, and the energy of the oxygen atom respectively. Because oxygen is a diatomic gas, which means that it occurs naturally as O$_2$, we calculate the total energy of one oxygen atom by dividing the total energy of O$_2$ by 2.
\subsection{Projected Density Of States}
We also calculated the Projected Density of States (PDOS). We plotted the PDOS of the $d$ orbital of the dopant and the $2p$ orbital of the oxygen. Before adsorption occurs, we plot the $d$ orbital of the dopant to identify if there were partially occupied d orbitals near the Fermi level. The Fermi level was set to 0 for all of our PDOS graphs. After adsorption, we plotted the 2p orbital from the oxygen as well as the d orbital of the dopant. We looked for strong hybridization, which occurs when the $2p$ and $d$ orbitals from the oxygen and dopant respectively rise and fall at the same time. 
\section{Results and Discussion}
First, we analyze the catalytic properties of pristine MoS$_2$ and Pd monolayers. We then present our results from the calculations pertaining to the doped MoS$_2$ and the doped MoS$_2$ with sulfur vacancies.
\subsection{Pristine Substrates}
\subsubsection{MoS$_2$}
\begin{figure}[hpt]
	\begin{subfigure}
	\centering
                \includegraphics[width=4cm]{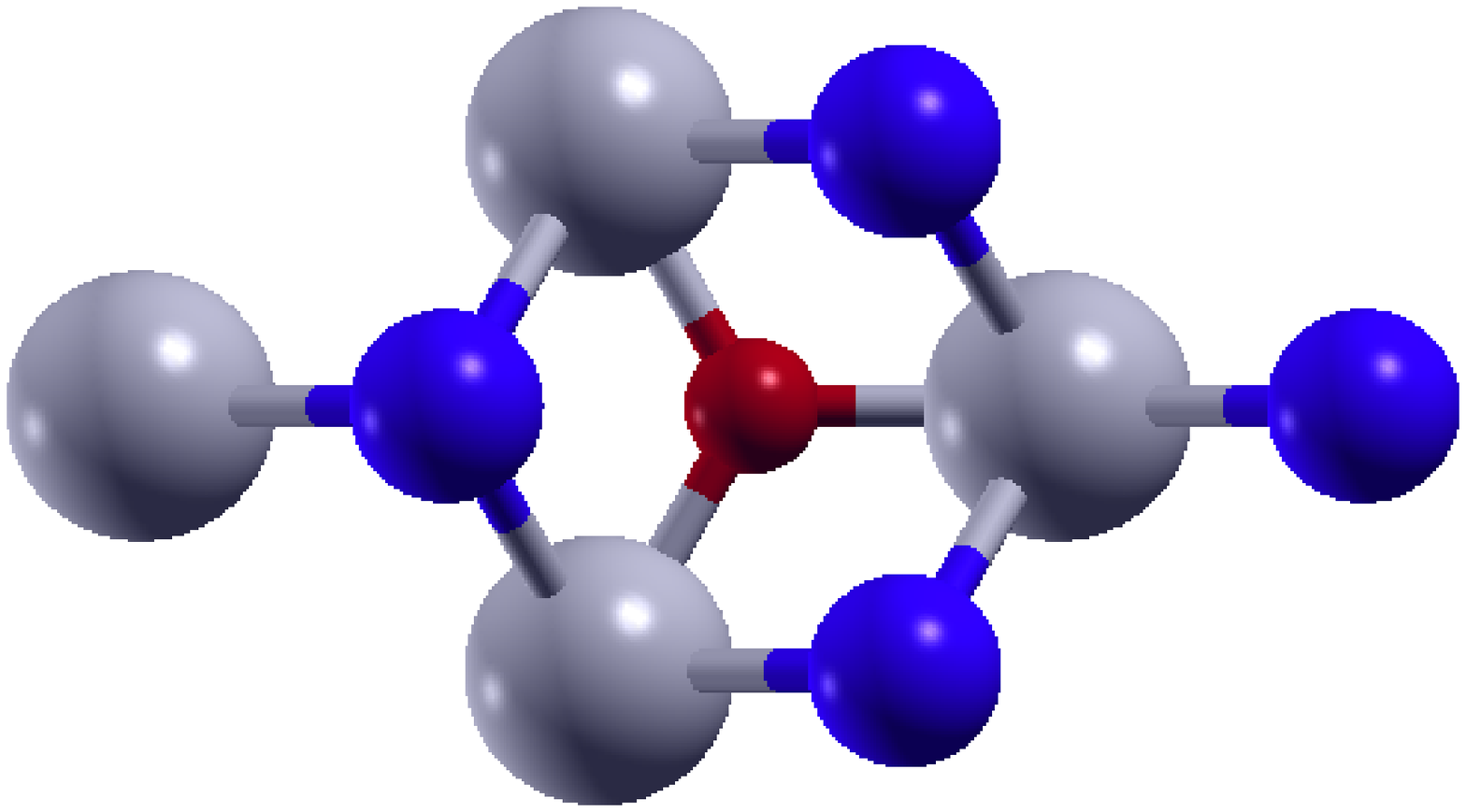}
                \includegraphics[width=4cm]{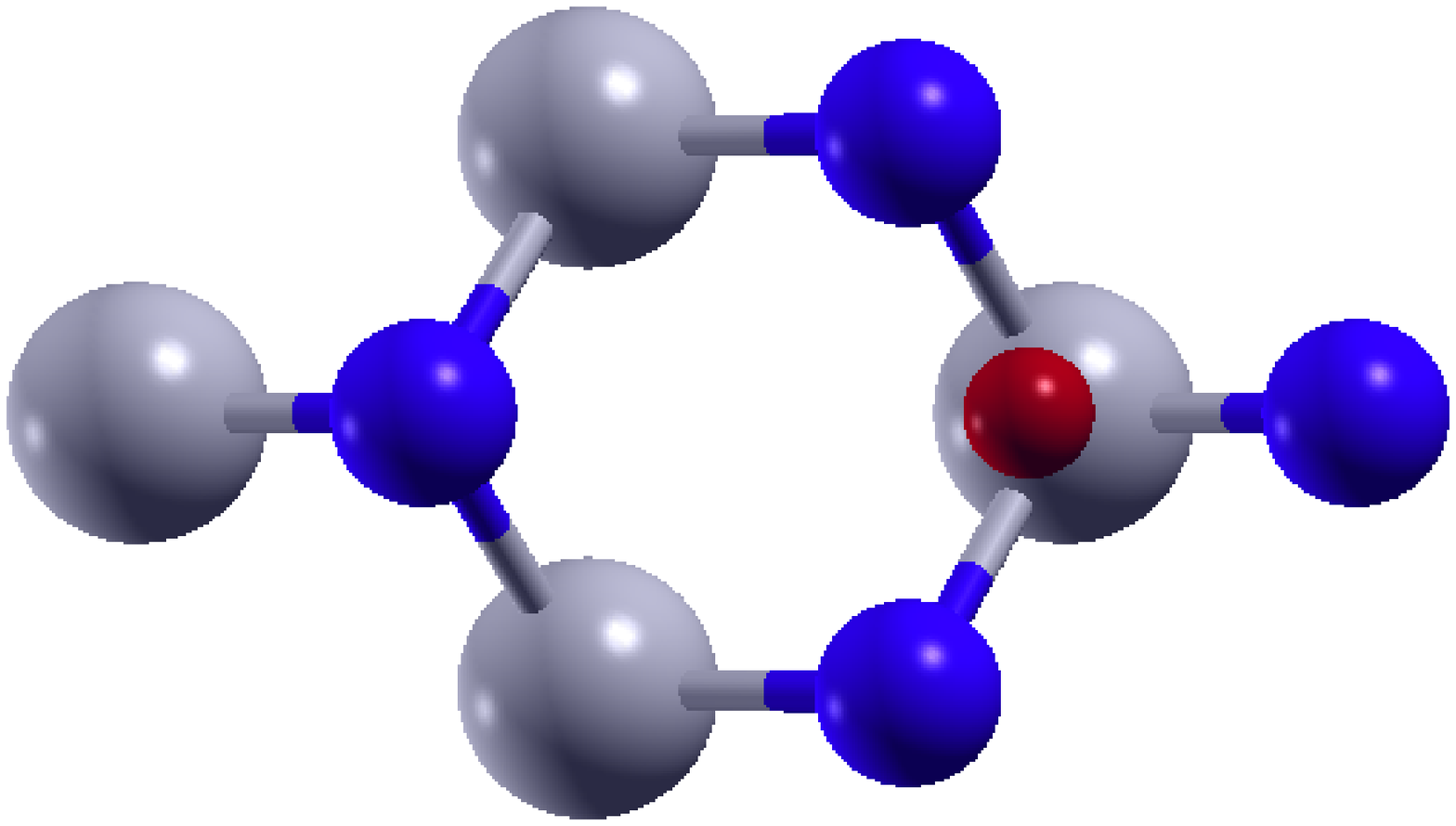}
		\includegraphics[width=4cm]{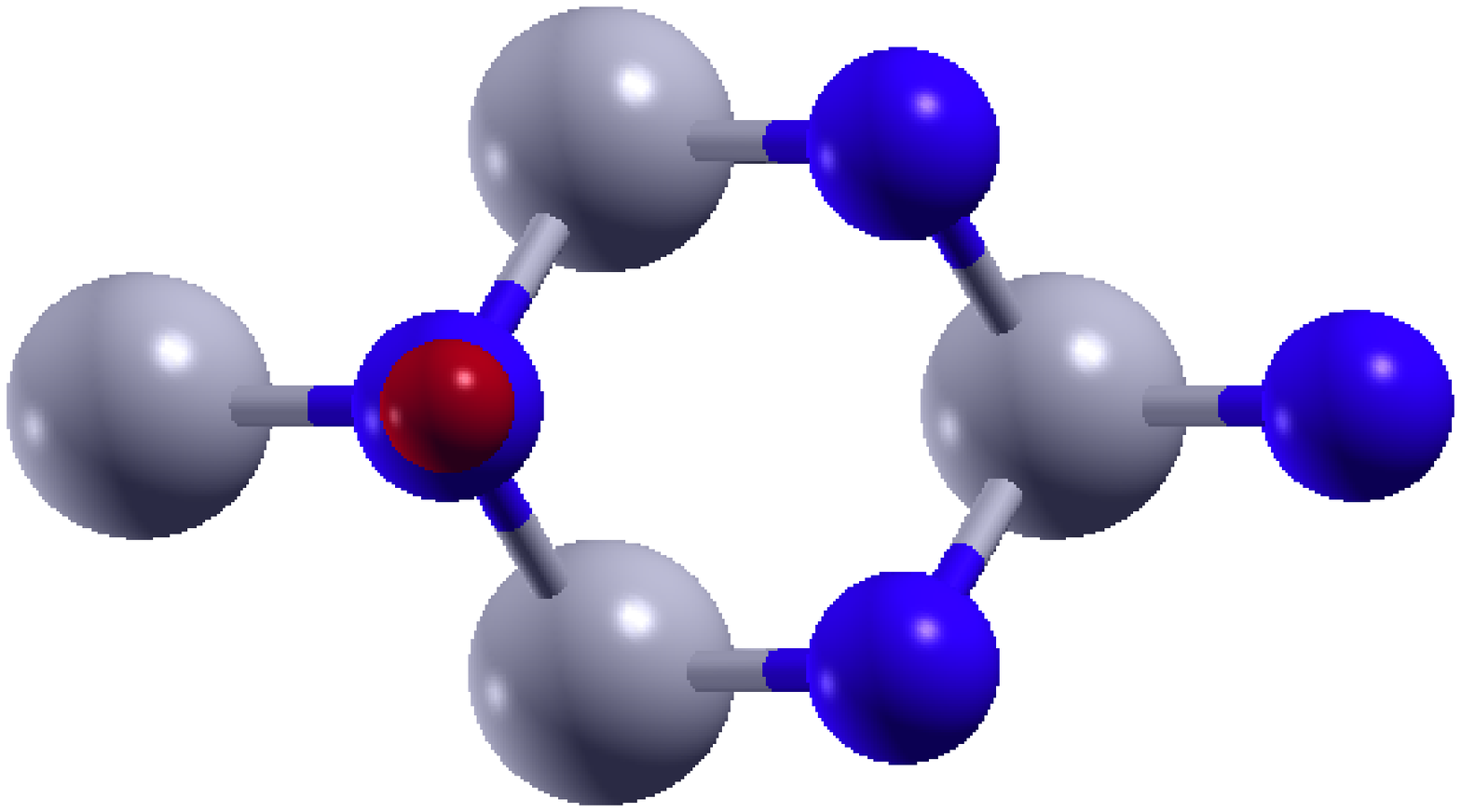}
\end{subfigure}
\begin{subfigure}
	(a)\includegraphics[width=4cm]{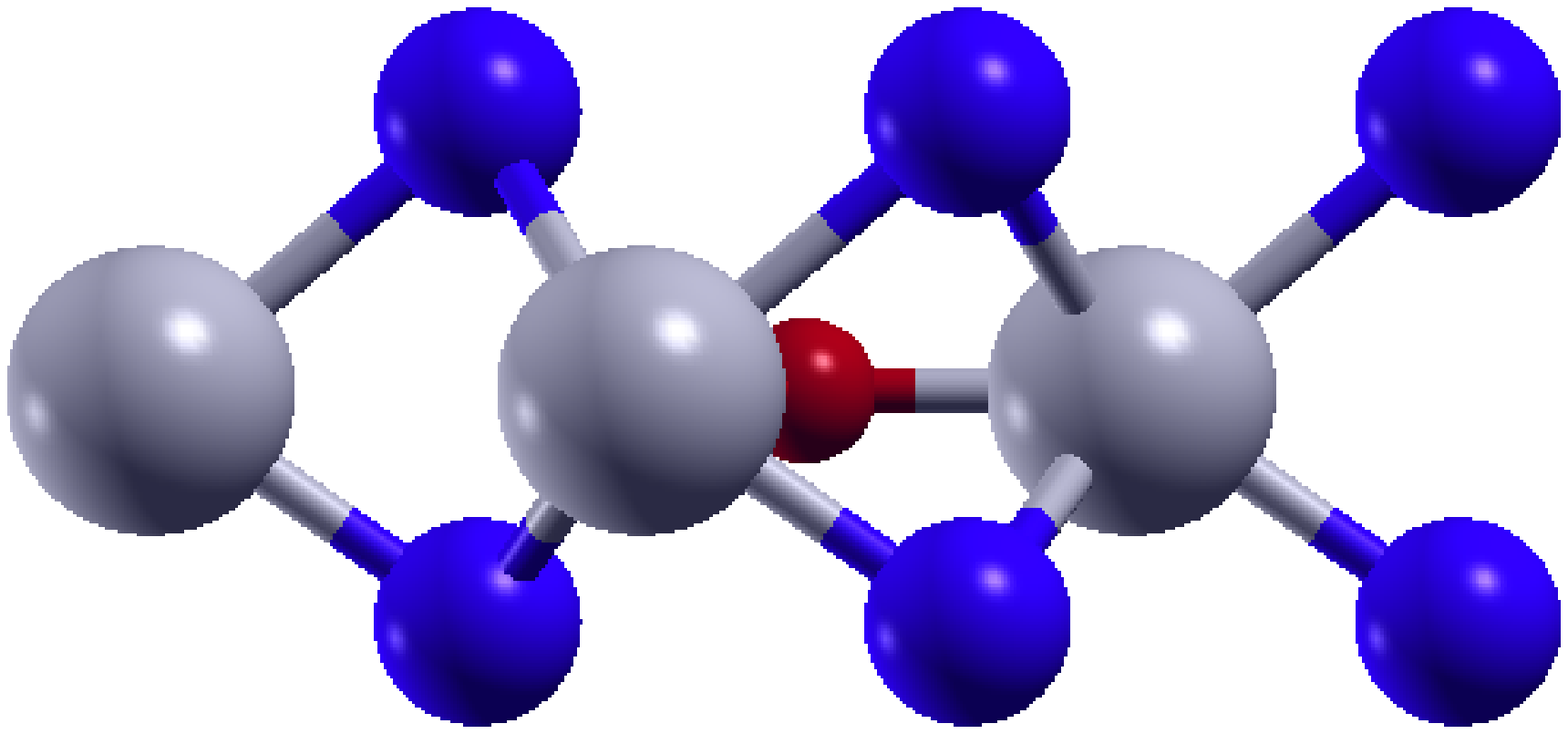}
	(b)\includegraphics[width=4cm]{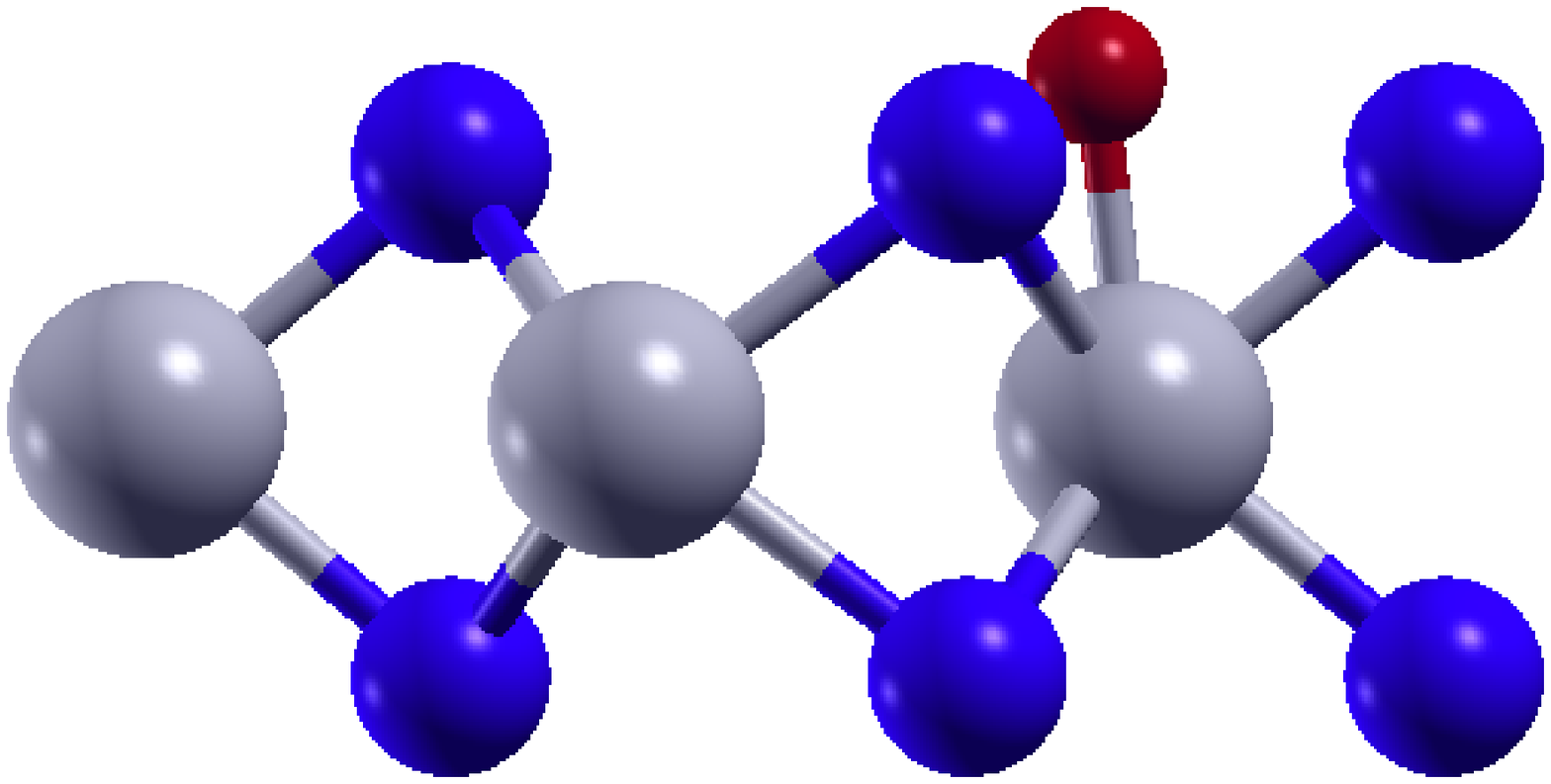}
	(c)\includegraphics[width=4cm]{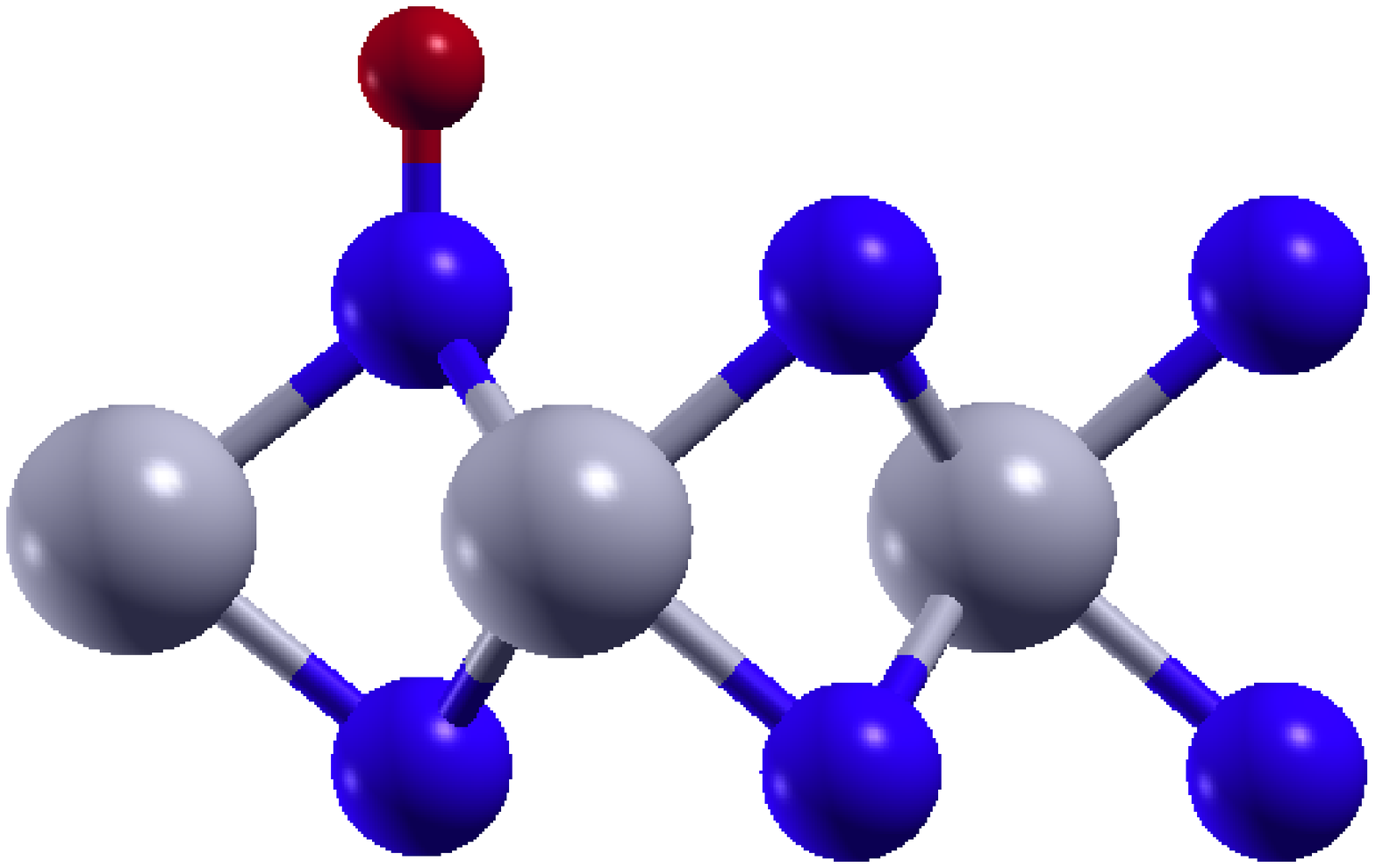}
\end{subfigure}	
	\caption{Atomic structures of MoS$_2$ after adsorbing oxygen. (a) represents the O adsorbed in the center of the cell. (b) represents the O adsorbed on the Mo. (c) represents the O adsorbed on the S atom. The first row represents the top view of the MoS$_2$ and the bottom row represents the side view of the MoS$_2$. The blue, gray and red atoms represent sulfur, molybdenum, and oxygen respectively.}
\label{Adsorptionsites}
\end{figure}

Figure \ref{Adsorptionsites} shows the atomic structure of MoS$_2$ before and after adsorbing oxygen. The figure shows the front and side views of the three adsorption sites. Figure \ref{Adsorptionsites} (a) shows the front and side views of the oxygen adsorbed in the center of the cell, (b) shows the front and side views of the oxygen adsorbed on the molybdenum atom, and (c) shows the front and side views of the oxygen adsorbed on top of the sulfur atom.
\par
We also measured the bond lengths and bond angles of the MoS$_2$. The relaxed lattice constants used was 5.98 Bohr, which is in good agreement with experimental values (5.97 Bohr)\cite{ataca2011comparative}. We measured the S-Mo-S bond angle to be 80.51 degrees, which has a 0.04\% error\cite{kadantsev2012electronic}.
\par
Using the relaxed lattice constants, total energy calculations were conducted. These energy values were then used in Equation 1 to calculate the adsorption energy of oxygen on the $2\times2\times1$ supercell of the material. A negative adsorption energy indicated that the material used was stable, as it did not gain energy after adsorbing an oxygen atom, and an exothermic reaction occurred. Conversely, positive adsorption energies indicated that an endothermic reaction occurred and the material was less stable as it gained heat. We looked for materials that had a negative adsorption energy with a large magnitude, which indicated that the bond between the oxygen and material was very strong. The adsorption energies calculated for the MoS$_2$ are shown in Table \ref{MoS2energies}. These values show that the best adsorption site for oxygen on a MoS$_2$ sheet is on top of the Molybdenum atom. When the oxygen is put on the sulfur, the oxygen is very unstable, and has a low adsorption energy at 0.825 eV. This is in agreement with past studies as other calculations show that the sulfur is not an active adsorption site, because the oxygen is very unstable when placed on the sulfur\cite{wang2015transition}.
\subsubsection{Palladium}
\begin{figure}[hpt]
        \subfigure[]{\includegraphics[width=6cm]{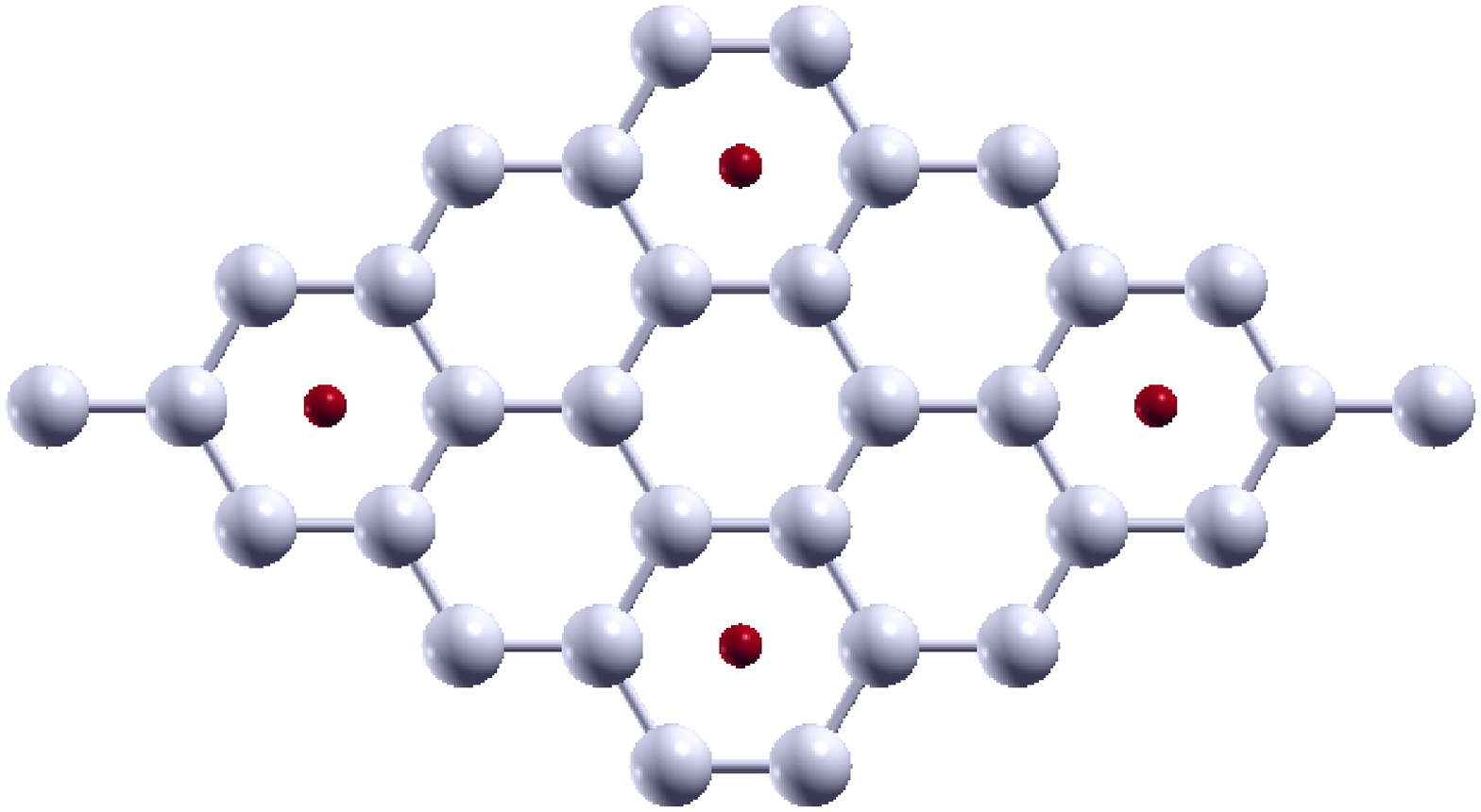}}
	\subfigure[]{\includegraphics[width=6cm]{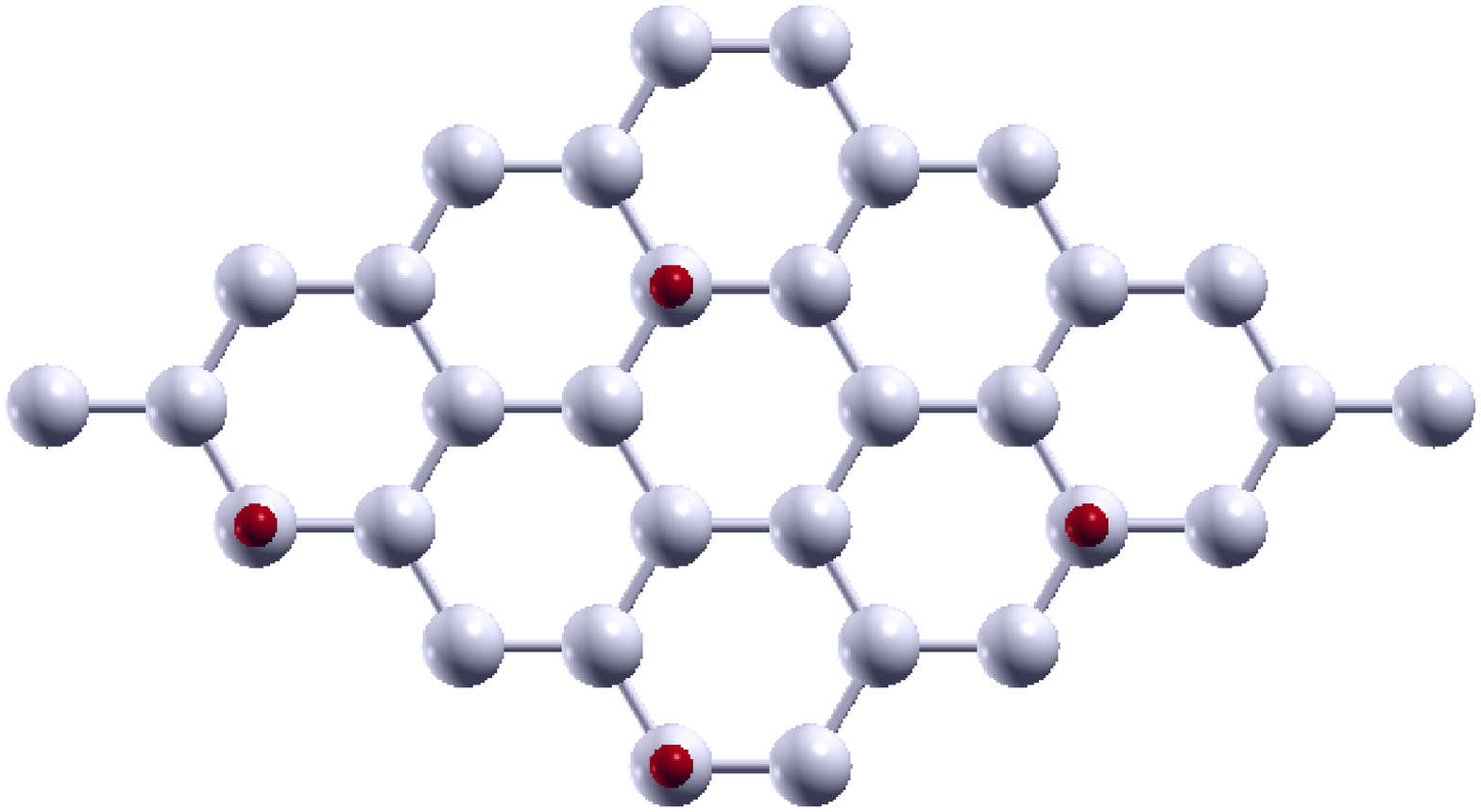}}
	\caption{The atomic structures of the two dimensional hexagonal monolayer of palladium: $($a$)$ shows the oxygen adsorbed in the center of the hexagonal monolayer; $($b$)$ shows the oxygen adsorbed on one of the Pd atoms. The gray atoms represent palladium, and the red atoms represent oxygen.}
	\label{Pdsites}
\end{figure}

	Figure \ref{Pdsites} shows the atomic structure of the 2$\times$2 hexagonal monolayer for Pd. We considered two adsorption sites for the palladium hexagonal monolayer. We placed the oxygen on one of the Pd atoms and in the middle of the hexagonal supercell. These adsorption sites are also shown in Figure \ref{Pdsites}. Figure \ref{Pdsites} (a) shows the top view of the Pd monolayer with the oxygen adsorbed in the center, and Figure \ref{Pdsites} (b) shows the top view of the Pd monolayer with the oxygen adsorbed on the Pd atom. We also calculated the lattice constants of the Pd monolayer. Our relaxed lattice parameter was 17.03 Bohr, which has a 0.1\% error when compared to the experimental values.
\par
	We found that our palladium monolayer had very similar adsorption energies to previous studies, because past research shows that the adsorption energy of one oxygen atom on the Pd atom is -2.21 eV\cite{shah2017oxygen}. Our study shows that the adsorption energy of O on Pd is -2.19 eV and -2.22 eV when the O was placed in the center of the cell and on the Pd atom respectively, which is in good agreement with Shah. et al\cite{shah2017oxygen}.
\subsection{Doping and Vacancies}
\begin{figure}[hpt]
	\begin{subfigure}
	\centering
                \includegraphics[width=4cm]{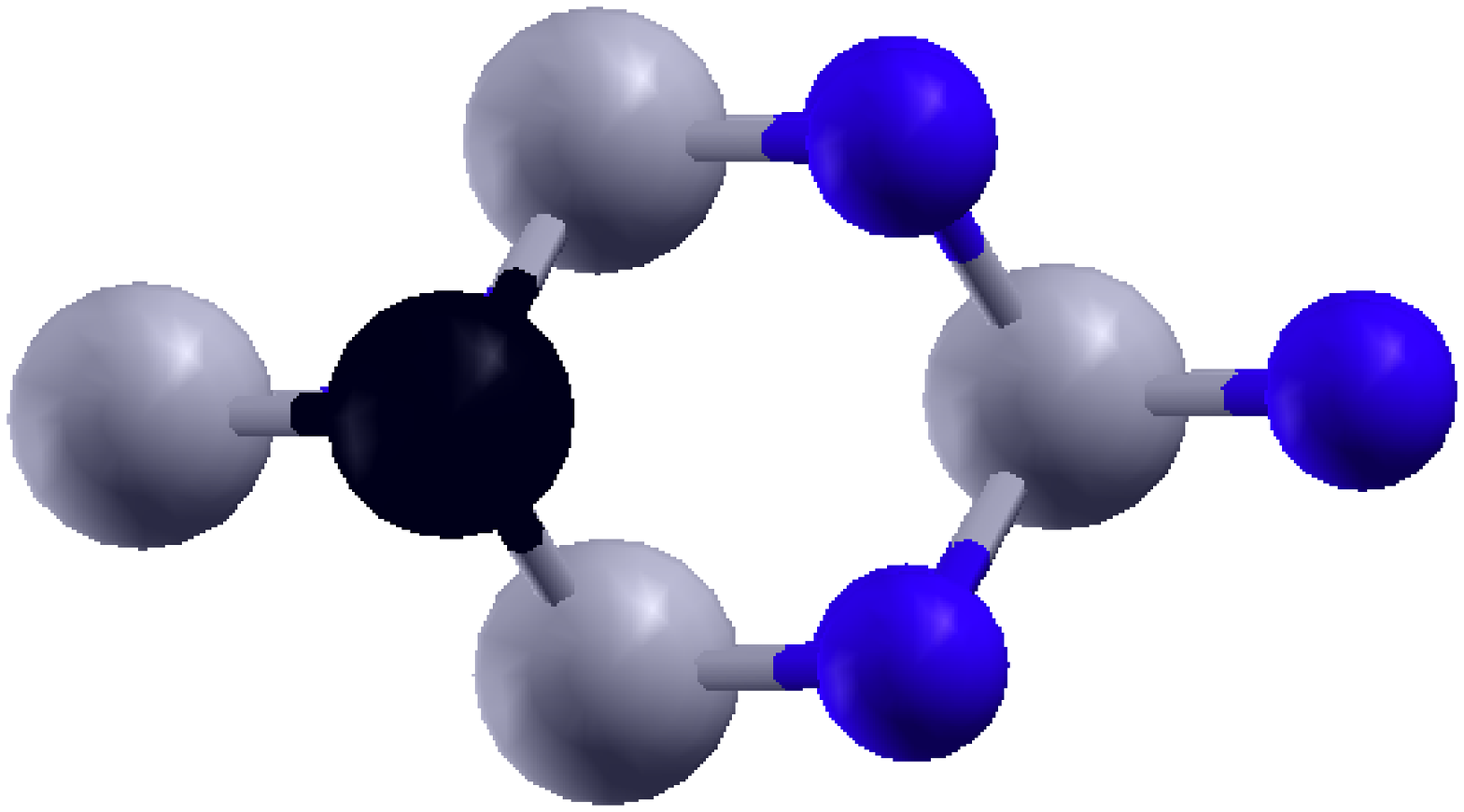}
                \includegraphics[width=4cm]{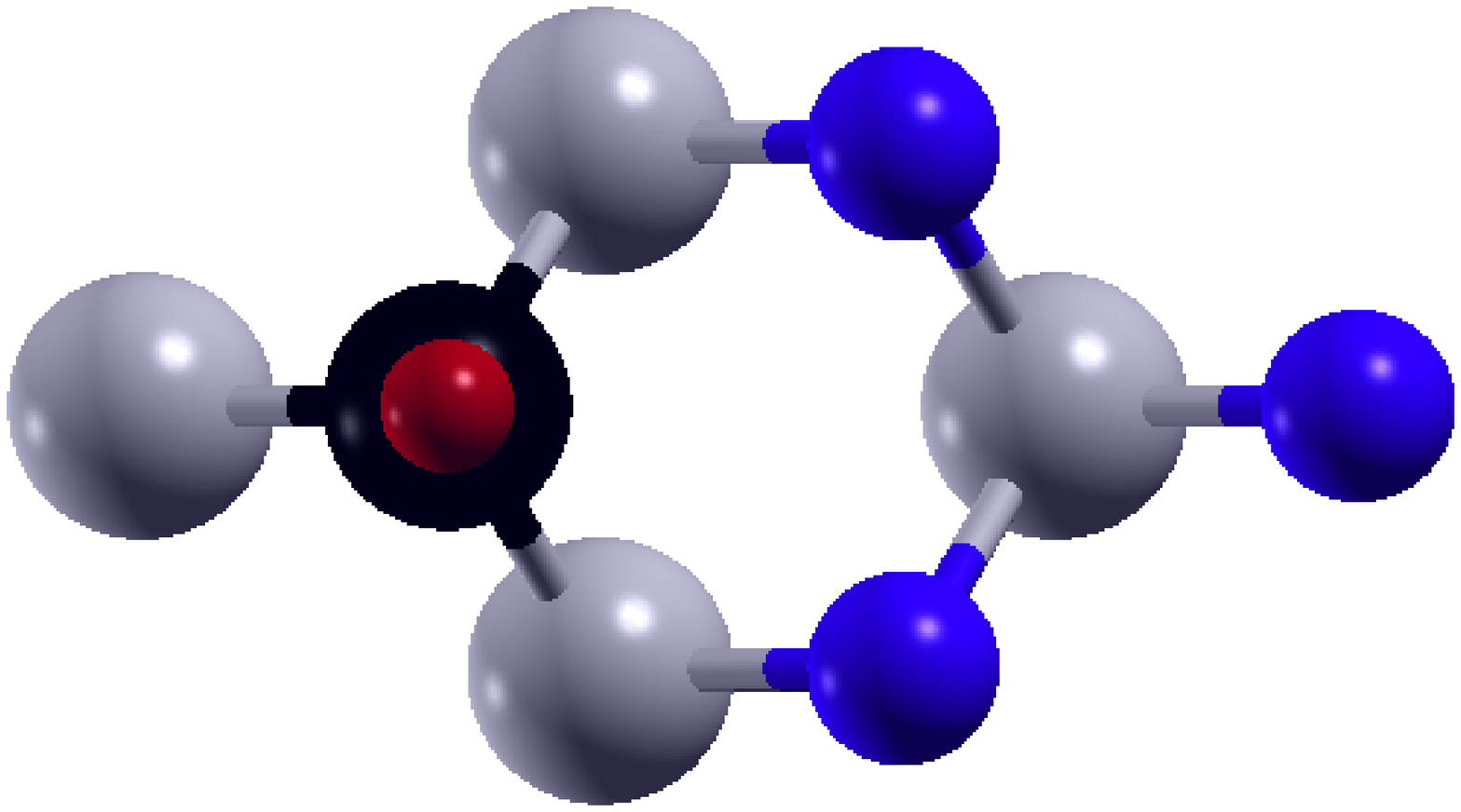}
		\includegraphics[width=4cm]{NiMoS2S.eps}
\end{subfigure}
\begin{subfigure}
	(a){\includegraphics[width=4cm]{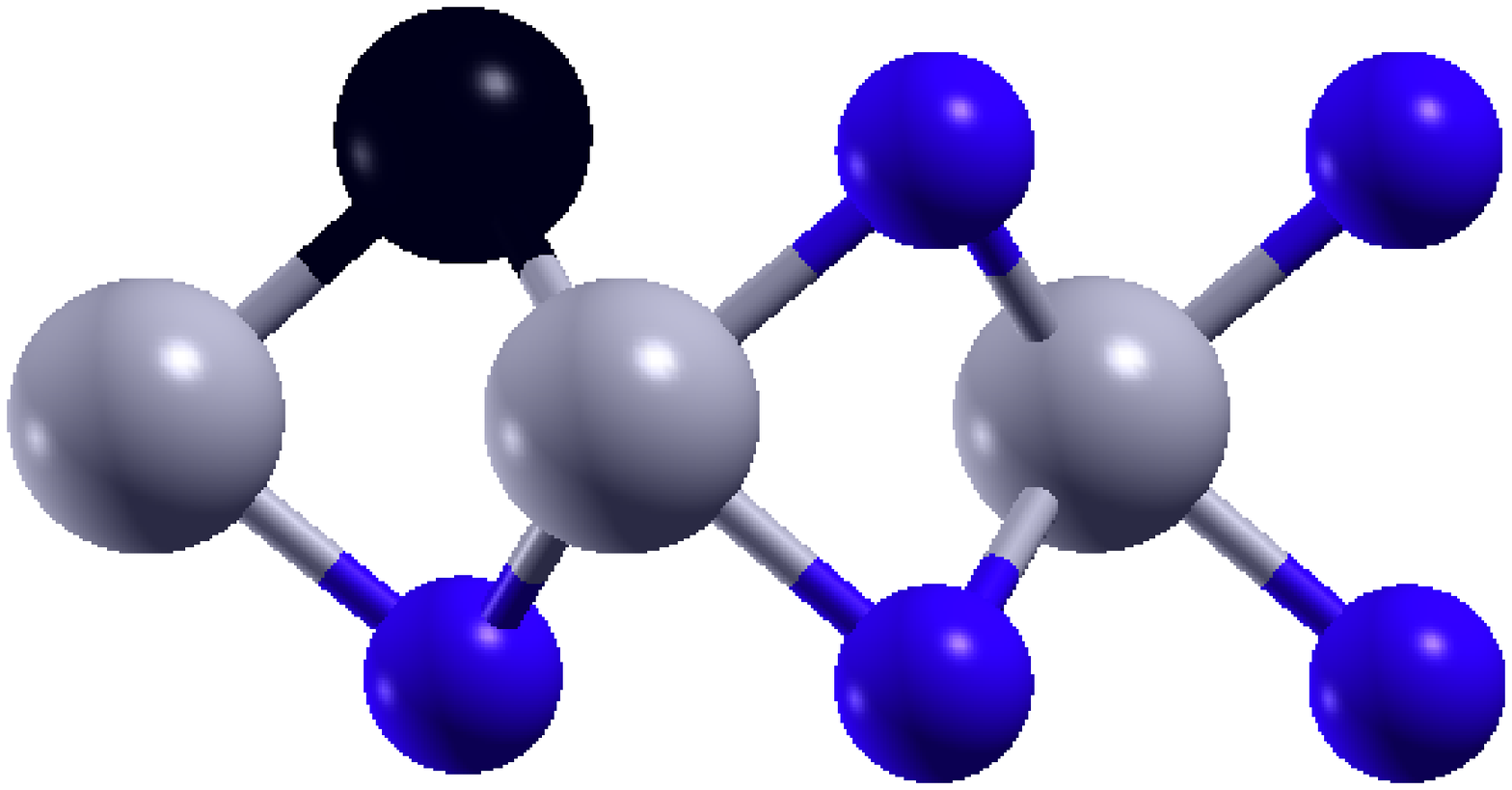}}
        (b){\includegraphics[width=4cm]{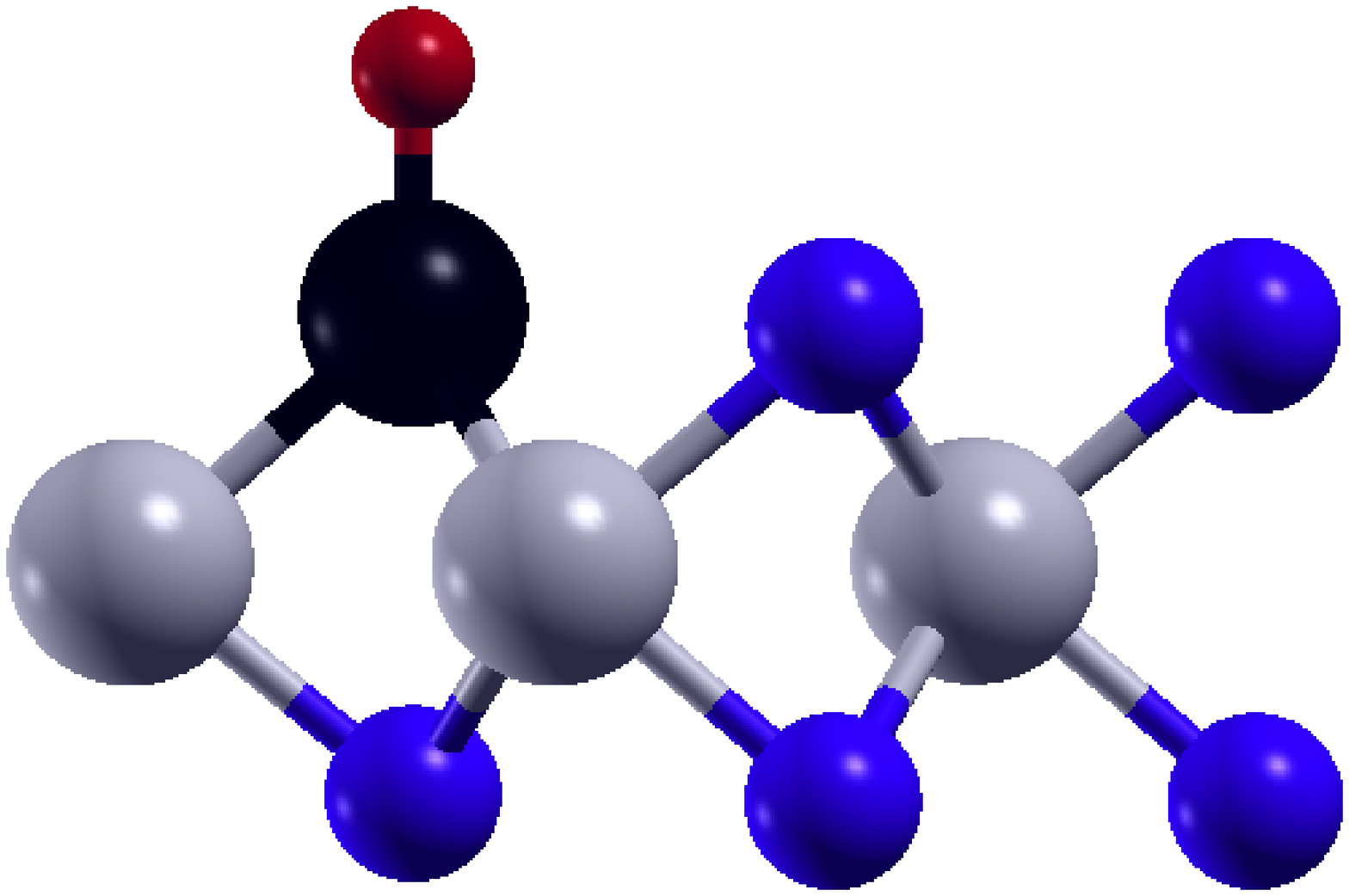}}
        (c){\includegraphics[width=4cm]{NiMoS2Sside.eps}}
\end{subfigure}	
	\caption{The atomic structures of (a) the doped MoS$_2$; (b) the doped MoS$_2$ after the O is adsorbed on the dopant; (c)  the doped MoS$_2$ after the O is adsorbed in a S vacancy. The first row represents the top view and the bottom row represents the side view. The gray, blue, black and red atoms represent molybdenum, sulfur, the transition metal, and oxygen respectively.}
        \label{dopantsites}
\end{figure}

After the MoS2 was substitutionally doped with the transition metals, two adsorption sites were considered. The doped MoS$_2$ is shown in Figure \ref{dopantsites} (a). We measure the adsorption energy of the oxygen on the MoS$_2$ after putting it on the dopant atom, as shown in Figure \ref{dopantsites} (b). Figure \ref{dopantsites} (c) shows the doped MoS$_2$ after the oxygen is adsorbed in a sulfur vacancy.
\par
After the oxygen molecule is adsorbed onto the doped MoS$_2$ sheets, the crystal structure changes significantly. Although the atomic radii of the dopants are larger than that of the sulfur atom, the TM dopants only slightly protrude from the molybdenum disulfide 0.04 \AA to 0.25 \AA. The average bond lengths between the Mo and dopant increased, after adsorption. These values are shown in Table \ref{delta}.
\subsection{Oxygen adsorption}
\subsubsection{O adsorption on Rh and Ir doped MoS$_2$}
From Table \ref{adsorption energy values}, it can be shown that the adsorption energy of O on Rh and Ir doped MoS$_2$ is very strong, with a magnitude of more than -4.5 eV. After a sulfur vacancy was introduced, the adsorption energy decreased below -6.8 eV. This is also shown in Table \ref{delta}, as the change in Mo-dopant bond length and the change in S-Mo-dopant angle of the Ir and Rh are greater than the change in bond length or bond angle for any other dopant. The strong interaction significantly activates the adsorbed molecule.
\par
\begin{figure}[htp]
               \subfigure[]{\includegraphics [width=6cm]{Irdos.eps}}
	       \subfigure[]{\includegraphics [width=6cm]{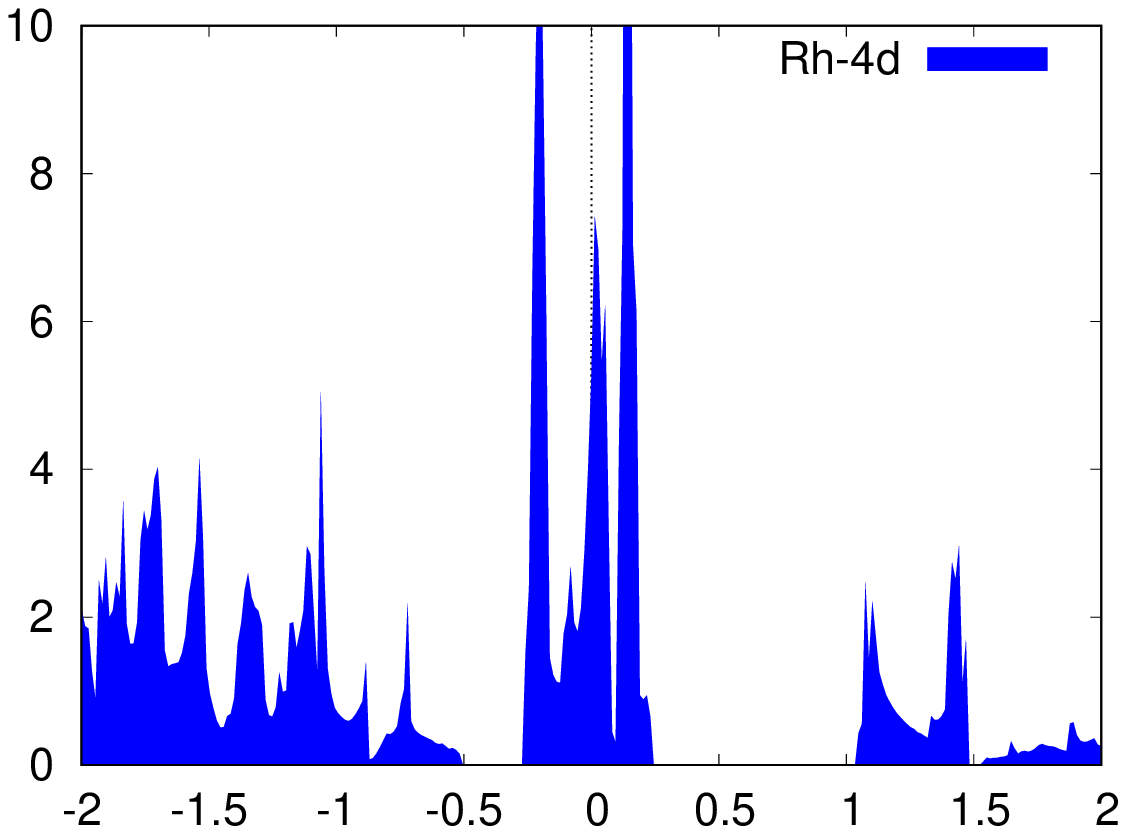}}
	       \subfigure[]{\includegraphics [width=6cm]{IrOdos.eps}}
	       \subfigure[]{\includegraphics [width=6cm]{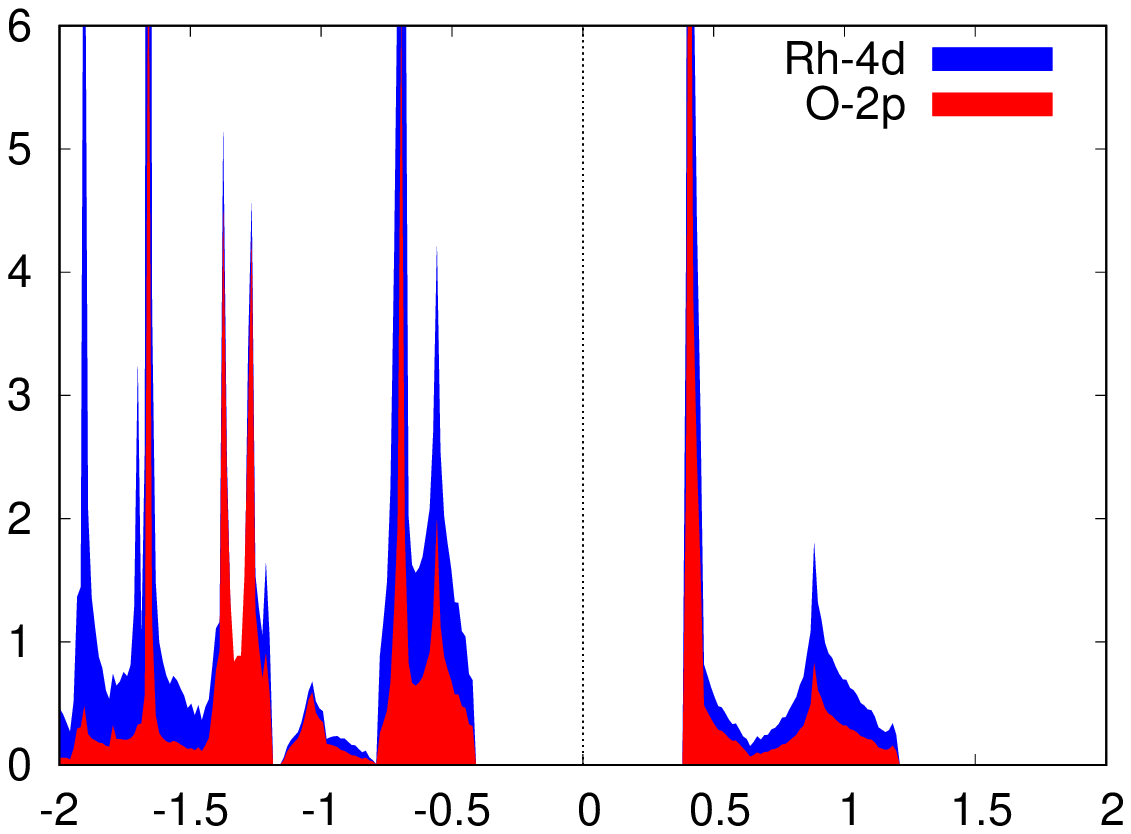}}
	\caption{PDOS for (a) Ir-5d, (b) Rh-4d, (c) Ir-5d and O-2p after adsorption, and (d) Rh-4d and O-2p after adsorption. The blue regions represent the d orbital of the dopant and the red area represents the 2p orbital of the oxygen.}
\label{pdosir}
\end{figure}

The analyses of electronic structures were performed in Figure \ref{pdosir}. We plotted the DOS of the 5d orbital for the Ir and the 4d orbital for the Rh. Previous studies suggest that the interaction of oxygen with the center of the transition metal dopants involves the electron transfer between each other\cite{ma2015co}. This means that partially occupied $d$ orbitals near the Fermi level are crucial to adsorb and activate O molecules as these orbitals facilitate the electron transfer from the substrate to the adsorbed O. Figure \ref{pdosir} shows that the Ir and Rh have partially occupied $d$ orbitals near the Fermi level, which shows that Ir and Rh have low adsorption energies.
\par
After adsorption, we also calculate the PDOS for the $d$ orbital of the dopant and the $2p$ orbital of the O. We looked for a strong hybridization in the DOS, which is shown by the simultaneous rise and fall of the $d$ and $p$ orbitals for the dopant and O respectively. This DOS is shown in Figure \ref{pdosir} (c) and \ref{pdosir} (d). Rh and Ir both have strong hybridization with the $2p$ orbital of the O. The projected density of states for the Ir and Rh intersect the $2p$ orbital of the O at approximately -1.5 eV. The Rh $4d$ orbital also intersects the O $2p$ orbital at 0.5 eV.
\par
These low adsorption energies are in agreement with previous studies as Ir and Rh doped MoS$_2$ were effective catalysts to adsorb O$_2$\cite{ma2016modulating}. Ma et al. also found that the Ir and Rh had partially occupied $d$ orbitals at the Fermi level, and that the crystal structure of Ir and Rh doped MoS$_2$ changed notably after adsorption ocurred\cite{ma2016modulating}. Fan et al.\cite{fan2017dft} also found that the Ir and Rh doped systems exhibited strong hybridization between the dopant and O as the orbitals occupied the same state and energy level. Past studies also indicate that the PDOS of these two dopants and the O strongly interact as the Projected Density of States intersects that of the O $2p$ orbital\cite{ma2015co}. This supports our data as the PDOS shows a strong hybridization between the dopant and the oxygen. 
\subsubsection{O adsorption on Fe and Co doped MoS$_2$}
\begin{figure}[htp]
	\subfigure[]{\includegraphics [width=6cm]{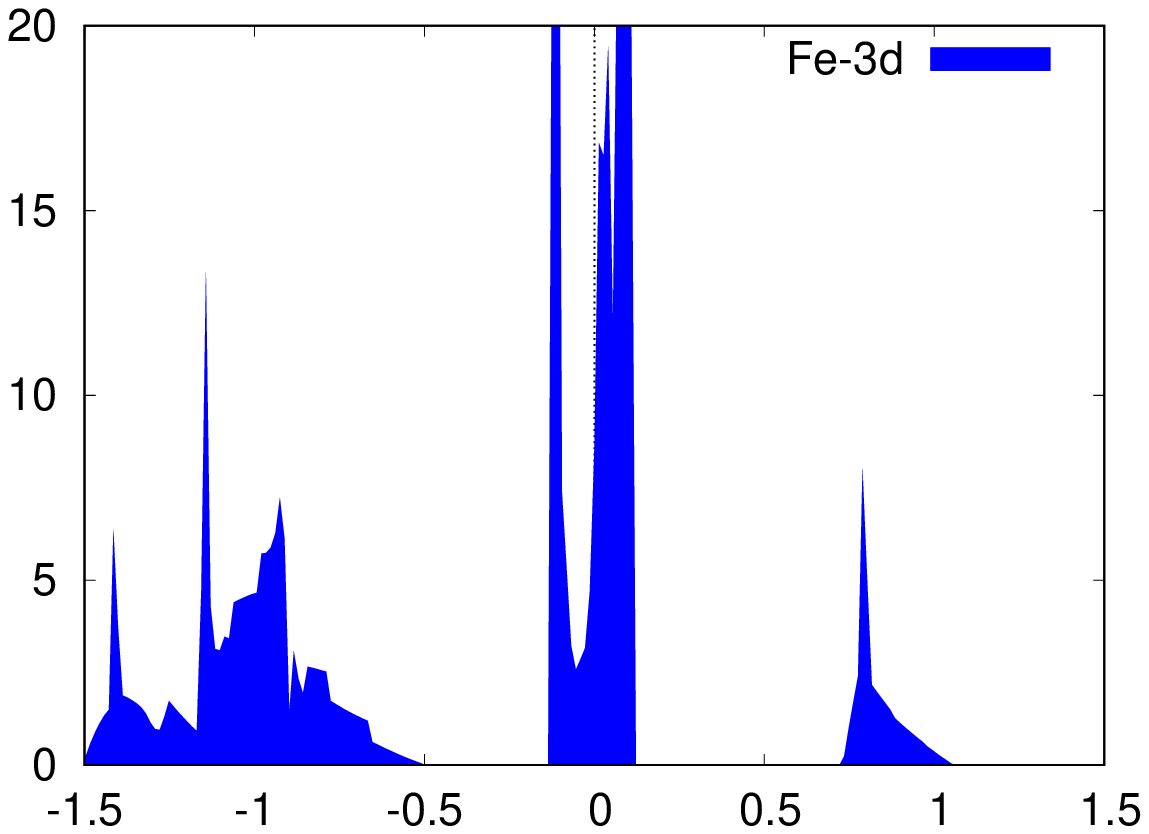}}
               \subfigure[]{\includegraphics [width=6cm]{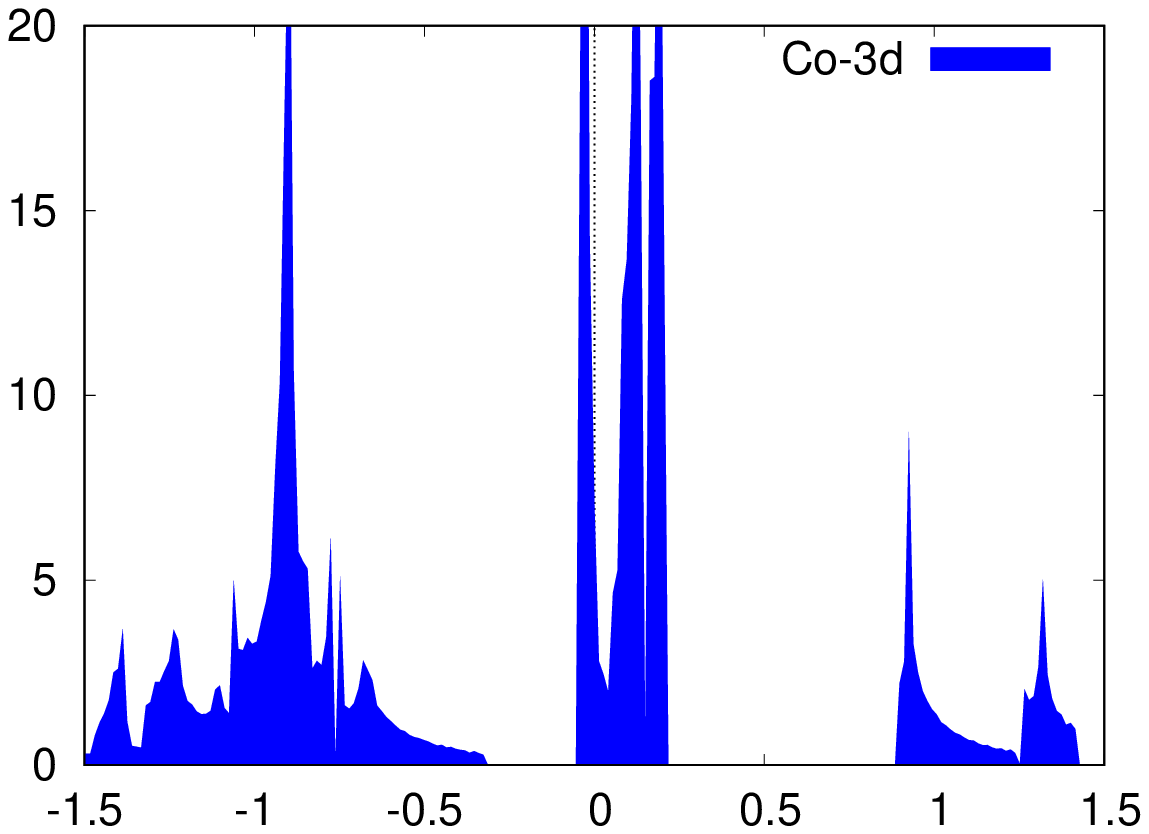}}
               \subfigure[]{\includegraphics [width=6cm]{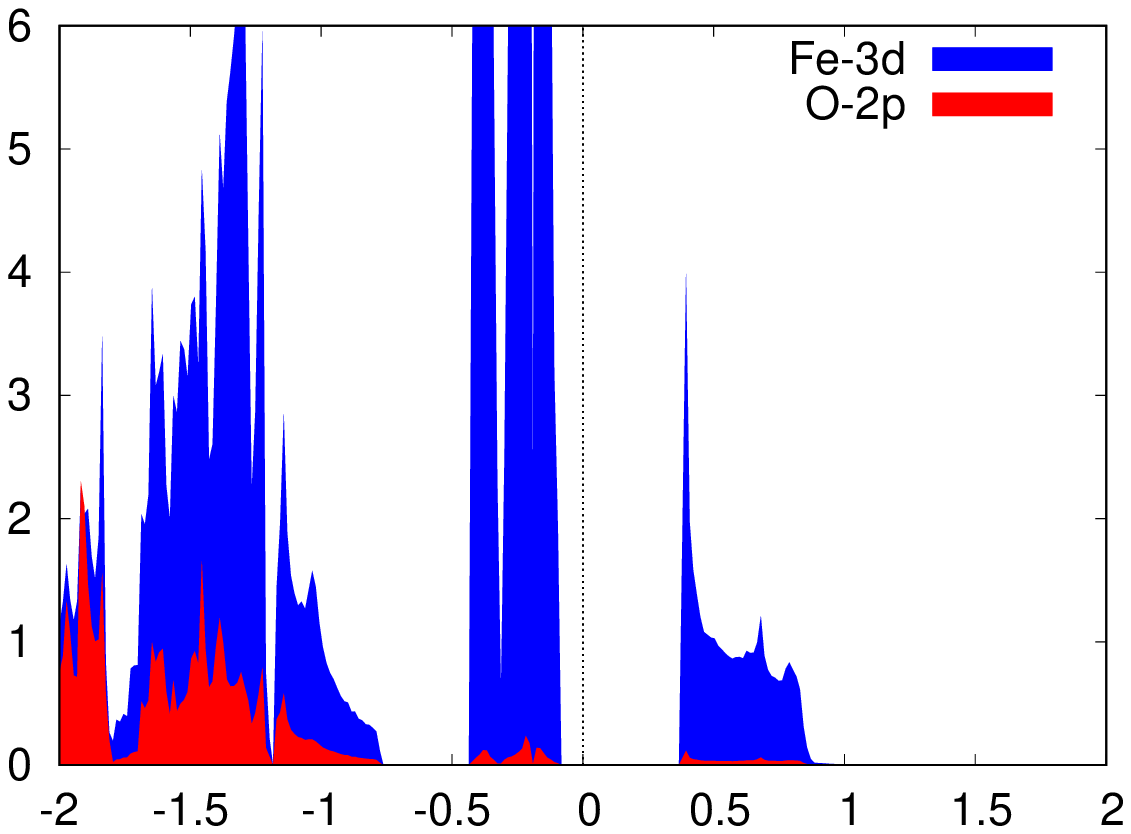}}
               \subfigure[]{\includegraphics [width=6cm]{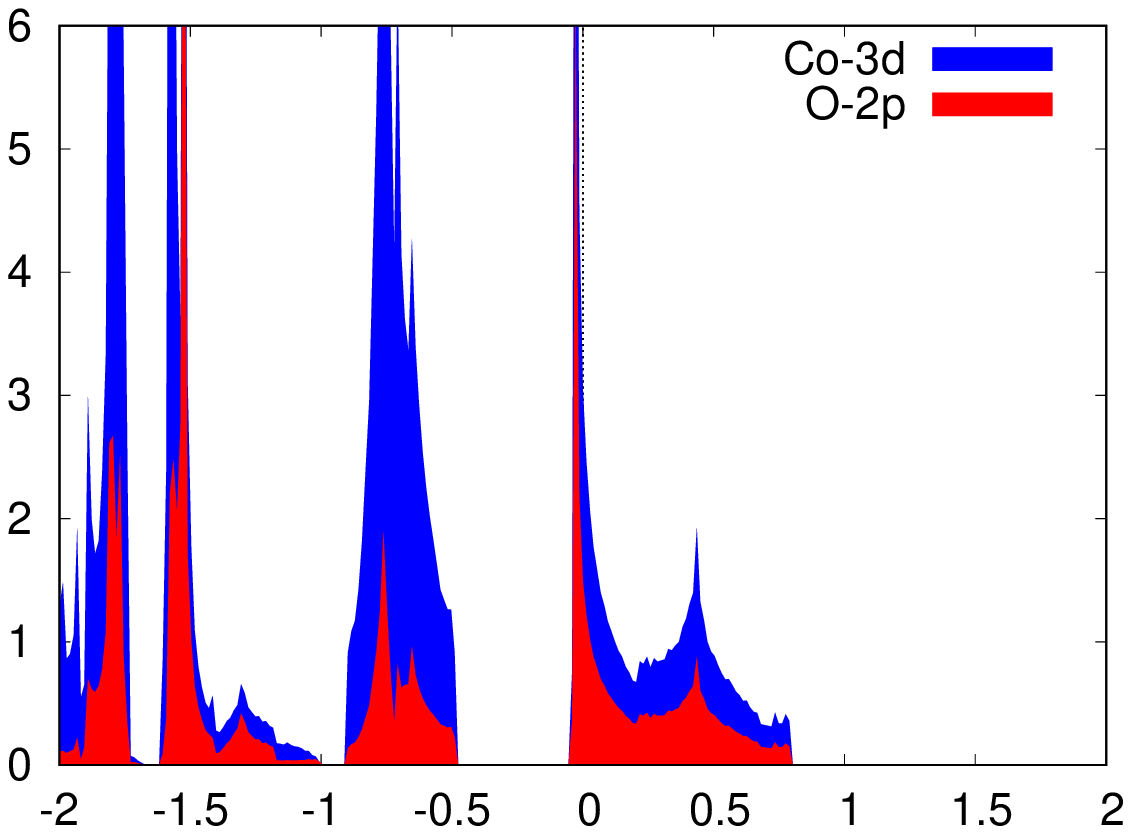}}
        \caption{PDOS for (a) Fe-3d, (b) Co-3d, (c) Fe-3d and O-2p after adsorption, and (d) Co-3d and O-2p after adsorption. The blue regions represent the 3d orbital of the dopant and the red area represents the 2p orbital of the oxygen.}
\label{Fedos}
\end{figure}

Iron and Cobalt were also very good catalysts as the adsorption energy of O on the Fe and Co-MoS$_2$ was greater than -4 eV, as shown in Table \ref{adsorption energy values}. This adsorption energy sharply decreased below -6.7 eV Both of these materials had partially occupied $d$ orbitals around the Fermi level, as shown in Figure \ref{Fedos}. These materials were slightly worse catalysts than Ir and Rh doped MoS$_2$ due to the weaker hybridization of the orbitals, shown in Figure \ref{Fedos} (c) and \ref{Fedos} (d). There is also no intersection between the 2p and 3d orbitals of the O and dopant respectively. Iron has very weak hybridization above -0.5 eV, and Co has weak hybridization between -2 and -1.5 eV. A strond bond still forms between these dopants and the O as the crystal structure changed significantly after adsorption ocurred. The change in the molybedum-dopant bond length was very large for both Co and Fe doped MoS$_2$ at 0.10 and 0.11 \AA respectively. The bond angle between the S-Mo-dopant also increased drastically at 3.72 and 3.39 degrees for Co and Fe separately. This shows that the O was strongly bonded to the doped MoS$_2$ because the bond lengths in the doped MoS$_2$ increased.
\par
Previous studies show that cobalt doped MoS$_2$ had a very low adsorption energy when adsorbing O$_2$\cite{xiao2015functional, fan2017dft}. Ma et al. found that the adsorption energy of oxygen on cobalt doped molybdenum disulfide was very similar to that of iridium or rhodium doped MoS$_2$. Our study shows that the adsorption of O on Co-MoS$_2$ is not as strong at the adsorpion of O on Ir or Rh-MoS$_2$, but this difference could have been caused by the difference in the adsorbate used. Ma et al.\cite{ma2016modulating} did find that the interaction between O$_2$ and Co was weaker than the interaction between the O$_2$ and the Rh or Ir. This supports the data in our PDOS as the co $3d$ and O $2p$ orbitals are not as strongly hybridized as the Rh or Ir $d$ orbitals and the O.
\par
Past research concerning iron doped MoS$_2$ has shown that Fe-MoS$_2$ is an effective catalyst to adsorb O$_2$, NO$_2$ and CO\cite{fan2017dft, ma2015co, chen}. Chen et al.\cite{chen} determined that the adsorption of gases on Fe doped MoS$_2$ is very strong, as the crystal structure changes significantly. This is in good agreement with our results in Table \ref{delta}. Fan et al. found that iron and cobalt exhibit similar catalytic properties, which validates the findings in our study. However, Fan et al. concluded that Fe and Co doped MoS$_2$ were more effective catalysts than Ir and Rh doped MoS$_2$. We can attribute this difference to the adsorbate used in the study. Ma et al.\cite{ma2015co} found that the adsorption of CO on Fe-doped MoS$_2$ did not produce strongly hybridized peaks even though Fe had a strong bond with the CO, which supports our data as we did not find strongly hybridized peaks.
\subsubsection{O adsorption on Ni doped MoS$_2$}
\begin{figure}[htp] 
         \subfigure[]{\includegraphics [width=6cm]{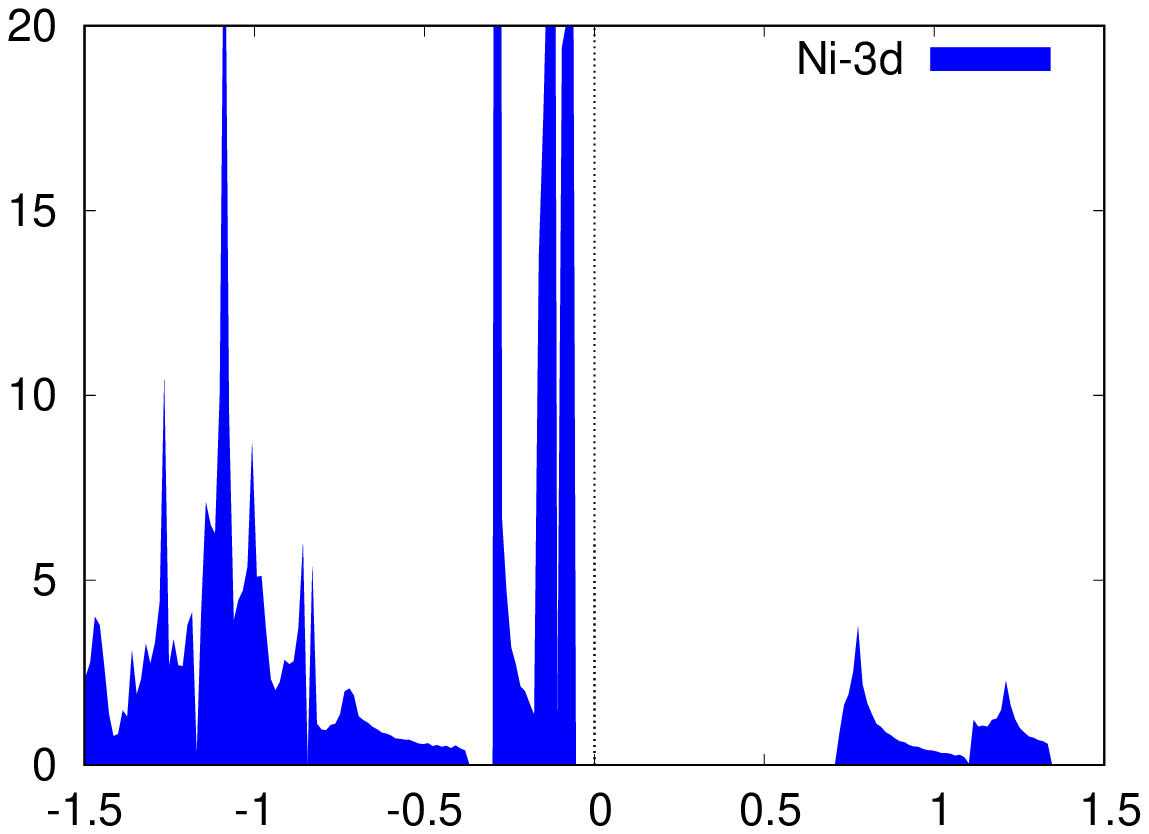}}
	       \subfigure[]{\includegraphics [width=6cm]{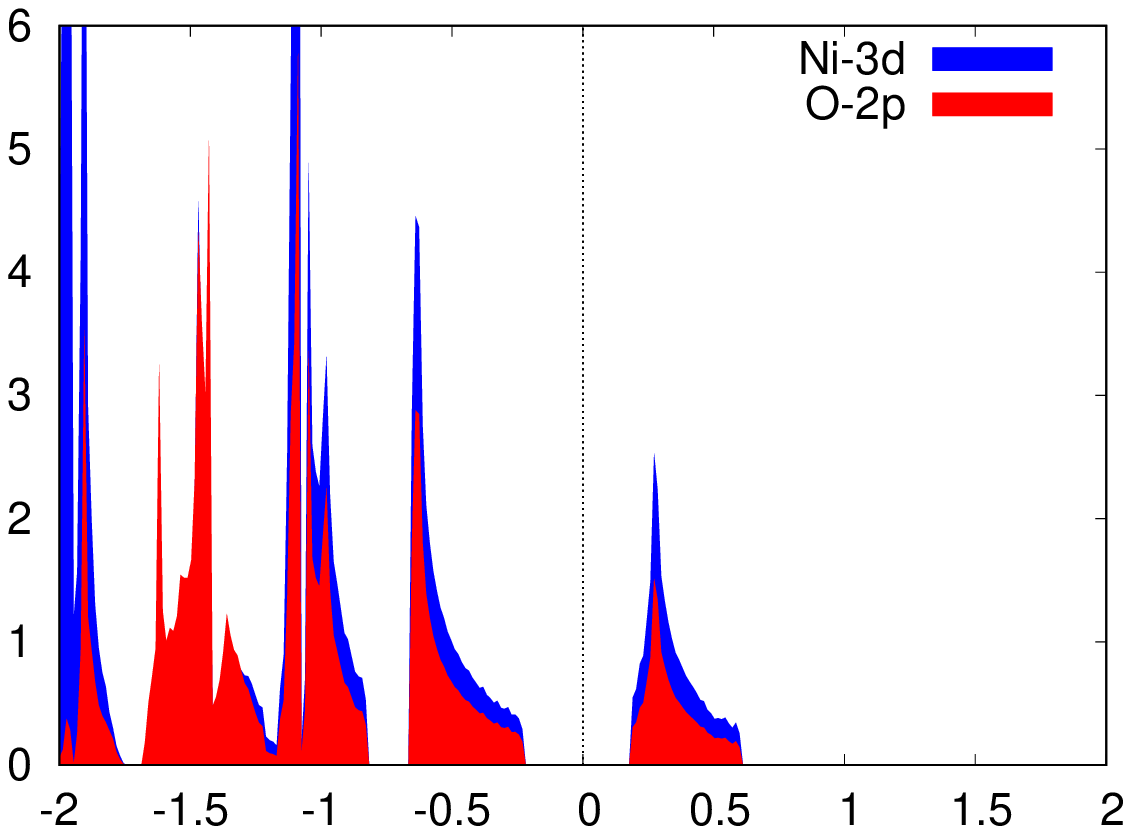}}
	\caption{PDOS for (a) Ni-3d, (b) Ni-3d and O-2p after adsorption. The blue and red regions represent the $3d$ orbital of the Ni and the $2p$ orbital of the O respectively.}
\label{Nidos}
\end{figure}

The adsorption energy of O on Ni is not as large as the adsorption energy of Co and Fe. With an adsorption energy of around -3.3 and -6.6 eV before and after adsorption respectively, the adsorption energy of O on Ni is slightly larger than that of Fe and Co. However, the PDOS does not have any peaks near the Fermi level, as shown in Figure \ref{Nidos} (a). The strong bond between the Ni is not caused by a partially occupied $d$ orbital, rather it is caused by the strong hybridization between the $2p$ and $3d$ orbitals from the O and Ni respectively, which is shown in Figure \ref{Nidos} (b). Furthermore, the p and d orbitals from the O and Ni intersect between -2 and -1.5. This indicates that a strong bond occurs, as the valence electrons of the Ni and O have the same energy level.
\par
The low adsorption energy of O on Ni agrees with previous studies conducted by Xiao et al\cite{xiao2015functional}. Xiao et al. found that the Ni did not have partially occupied $d$ orbitals, but concluded that the charge transfer between the Ni and O made Ni-MoS$_2$ a good catalyst. Xiao et al. did not study the hybridization between the O $2p$ and Ni $3d$ orbitals. However, Ma et al.\cite{ma2016modulating} found that Ni did not bond strongly to oxygen and did not hybridize strongly. This discrepancy can be attributed to the different asdorbate. The bond between the Ni-O is stronger than the bond between Ni and O$_2$ because all of the electrons in O$_2$ are in bonding orbitals and more energy is needed to excite the oxygen. Our adsorption energy of O on Ni-MoS$_2$ was -3.28 eV, which is in good agreement with previous data\cite{xiao2015functional} (-3.23 eV). Our value has a 1.5\% error when compared to previous results. 
\subsubsection{O adsorption on Ag and Cu doped MoS$_2$}
\begin{figure}[htp]
	\subfigure[]{\includegraphics [width=6cm]{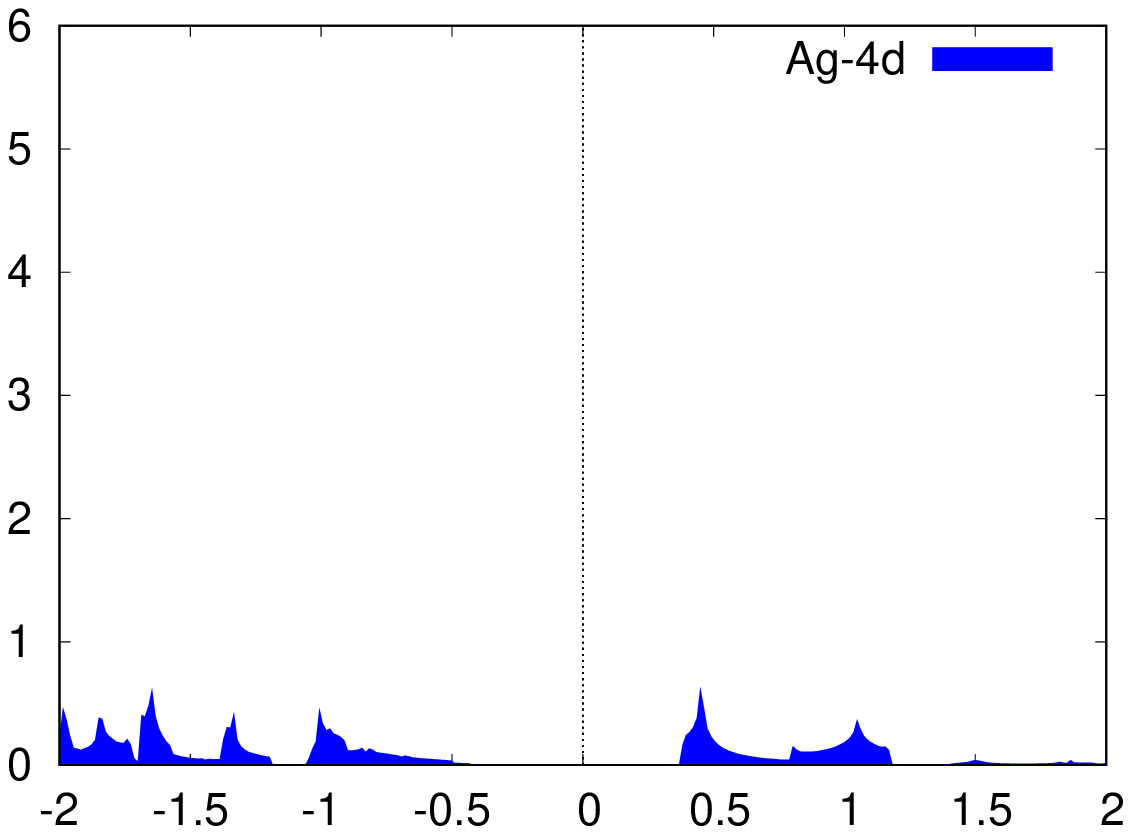}}
	\subfigure[]{\includegraphics [width=6cm]{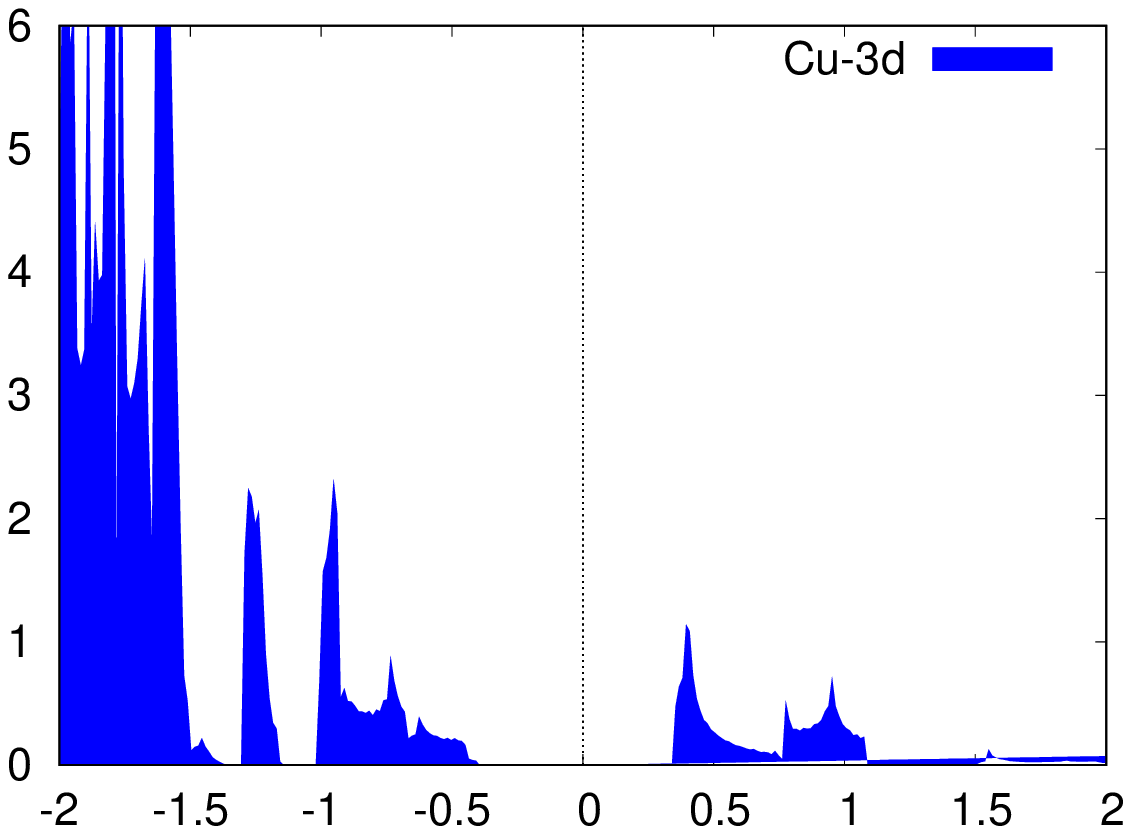}}
	\subfigure[]{\includegraphics [width=6cm]{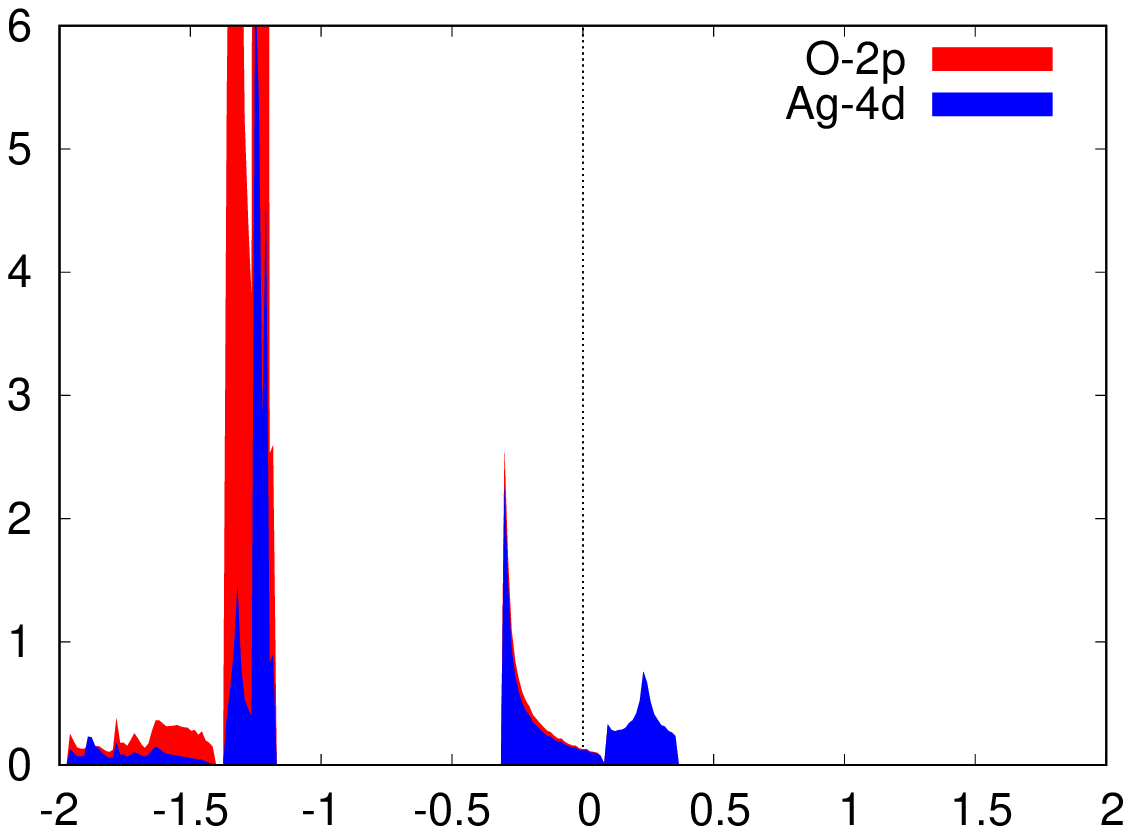}}
	\subfigure[]{\includegraphics [width=6cm]{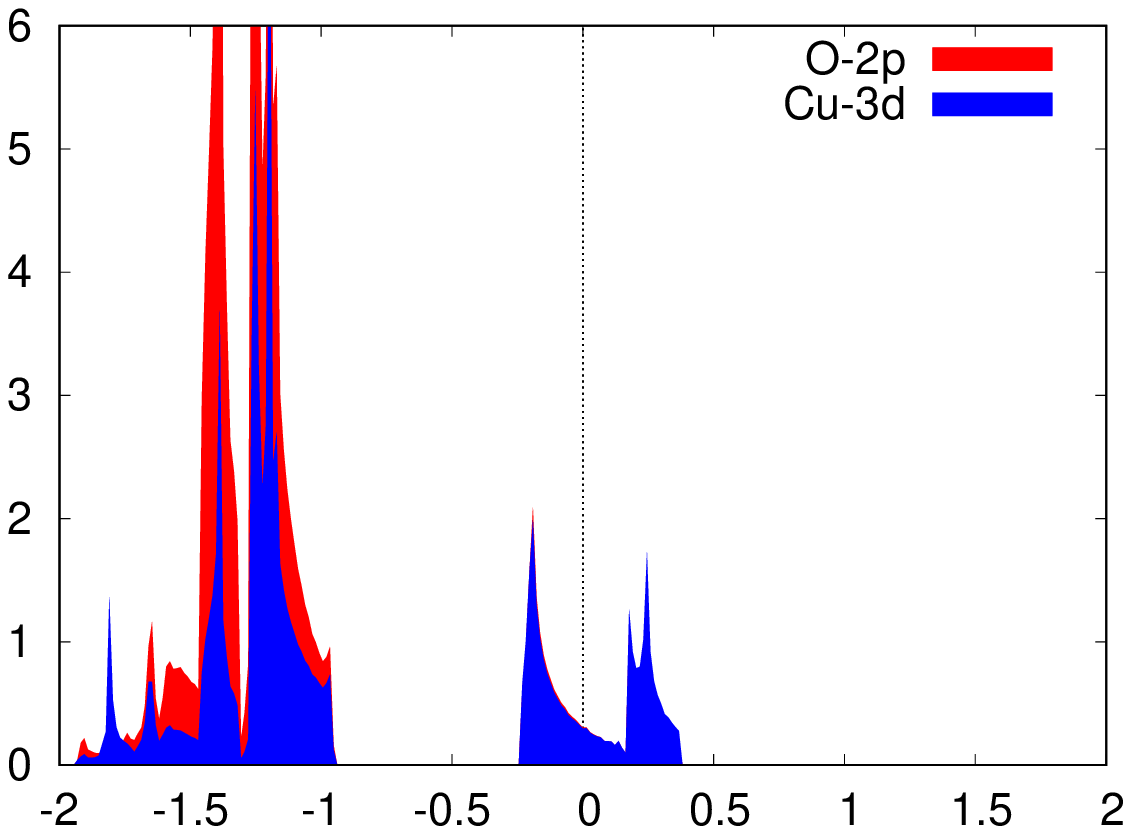}}
	\caption{PDOS for (a) Ag-4d, (b) Cu-3d, (c) Ag-4d and O-2p after adsorption, (d) Cu-3d and O-2p after adsorption. The blue regions represent the d orbital of the dopant and the red area represents the 2p orbital of the oxygen.}
\label{Agdos}
\end{figure}

The adsorption of O on Ag and Cu is relatively weak when compared to the other dopants. With adsorption energies of -1.18 and -2.01 eV respectively, the bond between these 2 metals and O is not as strong as the Ni-O bond. After sulfur vacancies are induced, this adsorption energy decreases to -6.28 and -6.48 eV for Ag and Cu respectively, as shown in Table \ref{adsorption energy values}. Neither Ag nor Cu have partially occupied $d$ orbitals near the Fermi level. This is shown in Figure \ref{Agdos} as the silver and copper doped MoS$_2$ do not have active $d$ orbitals. The adsorbed oxygen only modifies the crystal structure slightly, as the bond lengths only change by 0.04 and 0.01 \AA for Cu and Ag respectively. The change in the S-Mo-dopant bond angle is also very small at 1.09 and 1.62 degrees for Ag and Cu respectively. This small change in the crystal structure shows that the Cu and Ag do not strongly bond to the O. The crystal structure changes more for the Cu doped MoS$_2$ than for the Ag doped MoS$_2$, which is caused by the stronger hybridization between the Cu and O.
\par
The hybridization of Ag is also very weak, as shown in Figure \ref{Agdos} (c), which explains the very low adsorption energy of O on Ag doped MoS$_2$. There is no correlation between the peaks of the $2p$ orbital of the O and the $4d$ orbital of the Ag. The hybridization of the Cu is stronger, as shown in Figure \ref{Agdos} (d). There is a correlation between the peaks of the Cu and O for most values below 0 eV. However, no hybridization occurs above 0 eV.
\par
The low adsorption energy is in agreement with literature, as previous studies show that Ag-MoS$_2$ did not exhibit strong catalytic properties\cite{ma2016modulating, fan2017dft, chen}. However, past studies show that Cu-MoS$_2$ was a good catalyst to adsorb NO$_2$\cite{fan2017dft}. The disparity between the results can be attributed to a strong bond between the Cu and N. Chen et al.\cite{chen} found that the crystal structure of the Cu doped MoS$_2$ does not change as much as the Ni-MoS$_2$ after asdorption occurs. Chen at al.\cite{chen} also observed that the NO$_2$ did not contribute significantly to the DOS after adsorption, highlighting that there was little to no hybridization between the O $2p$ orbitals and the Cu $3d$ orbitals, which is in agreement with our data.
\section{Conclusion}
In this study, we use Density Functional Theory to investigate the effect of substitutional doping with TM atoms (Fe, Co, Ni, Cu, Rh, Ag, Ir) and sulfur vacancies on the electronic structure and catalytic properties of monolayer MoS$_2$. The Ir, Rh, Fe, Co and Ni-MoS$_2$ were better catalysts than the Pd monolayer as they had lower adsorption energies than the Pd. Our results suggest that Ir and Rh-MoS$_2$ have a strong bond with the oxygen, showing that they can break the carbon-oxygen bond in CO. This is mainly caused by the partially occupied $d$ orbital near the Fermi level, which is used to activate and adsorb O, and the strong hybridization between these dopants and O. A stronger hybridization indicates that the bond length decreases and the bond strength increases. We find that Fe and Co doped MoS$_2$ could also strongly adsorb O due to the partially occupied $d$ orbitals at the Fermi level. However, the hybridization between these dopants and O was not as strong as the hybridization between the Ir or Rh and O. Ni also had a low adsorption energy, but this was caused by the hybridization rather than the occupied $d$ orbital. Ag-MoS$_2$ and Cu-MoS$_2$ had the weakest adsorption energies as they had neither partially occupied $d$ orbitals nor a strong hybridization. These two dopants were not as effective as Pd. After sulfur vacancies were induced in the doped Mo$_2$ sheets, the adsorption energy of O on the MoS$_2$ decreased drastically, showing that the doped MoS$_2$ with S vacancies was a much better catalyst than the traditionally used palladium monolayer.
\par
Therefore, we conclude that a combination of doping and introducing sulfur vacancies could drastically improve the catalytic properties of MoS$_2$ monolayers. In the future, we would like to dope the MoS$_2$ with more transition metals that have similar electron configrations as Pd. We would also calculate the adsorption energy of various greenhouse gases on the MoS$_2$. This study shows that creating vacancies and doping can lead to the discovery of novel nanoparticle catalysts.
\newpage
\bibliography{sample_ref.bib}{}
\bibliographystyle{ieeetr}

\newpage
\begin{table}
\caption{Electron configuration and radius cutoff used to generate PAW pseudopotentials}
\begin{tabular}{|c|c|c|}
	\hline
        Material&Electron Configuration&PAW radius cutoff (Bohr)\\
\hline
\hline
	Oxygen&[He] 2s$^2$ 2p$^4$&1.4\\
\hline
	Sulfur&[Ne] 3s$^2$ 3p$^4$&1.9\\
\hline
	Iron&[Ne] 3s$^2$ 3p$^6$ 4s$^1$ 3d$^7$&2.1\\
\hline
	Cobalt&[Ne] 3s$^2$ 3p$^6$ 4s$^1$ 3d$^8$&2.1\\
\hline
	Nickel&[Ne] 3s$^2$ 3p$^6$ 4s$^2$ 3d$^8$&1.8\\
\hline
	Copper&[Ne] 3s$^2$ 3p$^6$ 4s$^1$ 3d$^{10}$&2.0\\
\hline
	Rhodium&[Ar 3d$^{10}$] 4s$^2$ 4p$^6$ 5s$^1$ 4d$^8$&2.2\\
\hline
	Palladium&[Ar 3d$^{10}$] 4s$^2$ 4p$^6$ 5s$^1$ 4d$^9$&2.5\\
\hline
	Silver&[Kr] 5s$^1$ 4d$^{10}$&2.5\\
\hline
	Iridium&[Kr 5s$_2$ 4d$_{10}$ 4f$_{14}$] 5p$_6$ 6s$_1$ 5d$_8$&2.5\\
\hline
\end{tabular}
\label{valence}
\end{table}

\begin{table}[hpt]
\caption{Crystal structure and converged kinetic energy cutoff for the dopants}
	\begin{tabular}{|c|c|c|}
\hline
		Material&Crystal Structure&Energy Cutoff (Ha)\\
\hline
\hline
		Fe&BCC&23\\
		\hline
		Co&Simple Hexagonal&24\\
		\hline
		Ni&FCC&40\\
		\hline
		Cu&FCC&22\\
		\hline
		Rh&FCC&16\\
		\hline
		Ag&FCC&16\\
		\hline
		Ir&FCC&19\\
		\hline
\end{tabular}
\label{dopant ecut}
\end{table}

\begin{table}[hpt]
\caption{Adsorption energies of O on pristine MoS$_2$}
\begin{tabular}{|c|c|}
       \hline
	Location&Adsorption Energy (eV)\\
\hline
\hline
        Center&-1.105\\%
	\hline
        Oxygen on Mo atom&-2.58\\%
       \hline
	Oxygen on S atom&0.825\\%
       \hline
	Sulfur Vacancy&-2.241\\%
	\hline
\end{tabular}
\label{MoS2energies}
\end{table}

\begin{table}[hpt]
\caption{Increase in dopant-Mo-S bond angle and bond length between Mo and dopant after adsorption.}
	\begin{tabular}{|c|c|c|}
\hline
		Dopant&Increase in dopant-Mo-S bond angle&Increase in Mo-dopant bond length (\AA)\\
\hline
\hline
		Ir&4.02$^\circ$&0.21\\%
		\hline
		Rh&3.83$^\circ$&0.19\\%
		\hline
		Co&3.72$^\circ$&0.11\\%
		\hline
		Ag&1.09$^\circ$&0.01\\%
		\hline
		Cu&1.62$^\circ$&0.04\\%
		\hline
		Fe&3.39$^\circ$&0.10\\%
		\hline
		Ni&3.82$^\circ$&0.19\\%
		\hline
\end{tabular}
\label{delta}
\end{table} 
\newpage
\begin{table}[hpt]
\caption{Adsorption energies of oxygen on the doped MoS$_2$}
\begin{tabular}{|c|c|c|}
\hline
        Dopant&Oxygen Location&Oxygen Adsorption Energy (eV)\\
\hline
\hline
        Ag&On Ag&-1.18\\
\hline
        Ag&Sulfur vacancy&-6.28\\
\hline
	Cu&On Cu&-2.01\\
	\hline
	Cu&Sulfur vacancy&-6.48\\
	\hline
        Co&On Co&-4.02\\
\hline
        Co&Sulfur vacancy&-6.72\\%
\hline
        Ni&On Ni&-3.28\\
\hline
        Ni&Sulfur vacancy&-6.58\\
\hline
        Fe&On Fe&-4.05\\
\hline
        Fe&Sulfur vacancy&-6.74\\
\hline
        Ir&On Ir&-4.58\\%
\hline
        Ir&Sulfur vacancy&-6.91\\%
\hline
        Rh&On Rh&-4.52\\%
\hline
        Rh&Sulfur vacancy&-6.88\\%
\hline
\end{tabular}
\label{adsorption energy values}
\end{table}

\newpage
\newpage
\end{document}